\documentclass[journal, onecolumn,12pt, draftclsnofoot]{IEEEtran}

\usepackage{mathrsfs}

\usepackage[noadjust]{cite}
\usepackage{graphicx,color,overpic,psfrag}
\usepackage{amsmath, amssymb}
\usepackage{latexsym}
\usepackage{bm}
\usepackage{amssymb}
\usepackage{cases}
\usepackage{array}
\usepackage{fancyhdr}
\usepackage{setspace}

\ifCLASSOPTIONcompsoc
\usepackage[caption=false,font=normalsize,labelfont=sf,textfont=sf]{subfig}
\else
\usepackage[caption=false,font=footnotesize]{subfig}
\fi

\usepackage{url}
\usepackage{algpseudocode}
\usepackage{algorithm}
\usepackage{blkarray}
\usepackage{booktabs}

\usepackage{multirow}
\usepackage{dsfont}
\usepackage{tabularx}
\usepackage[table]{xcolor}

\usepackage{amsfonts}

\usepackage{letltxmacro}

\graphicspath{{figure/}}

\newcommand{\ra}[1]{\renewcommand{\arraystretch}{#1}}

\newtheorem{prop}{Proposition}

\newcommand{\figref}[1]{Fig. \ref{#1}}
\newcommand{\tabref}[1]{Table \ref{#1}}

\newcommand{\secref}[1]{Section \ref{#1}}

\newcommand{\propref}[1]{Proposition \ref{#1}}

\newcommand{\Exp}{{\mathsf{E}}}
\newcommand{\expect}[1]{\Exp\left\{#1\right\}}

\newcommand{\tr}[1]{\mathsf{tr}\left\{#1\right\}}
\newcommand{\diag}[1]{\mathsf{diag}\left\{#1\right\}}

\newcommand{\abs}[1]{\left|#1\right|}

\newcommand{\oneon}[1]{\frac{1}{#1}}
\newcommand{\zeroldots}[1]{0,1,\ldots,#1}
\newcommand{\intd}{\mathrm{d}}
\newcommand{\intdx}[1]{\intd #1}

\newcommand{\GCN}[2]{\mathcal{CN}\left( #1 , #2\right) }
\newcommand{\argmax}[1]{\mathop{\arg\max}\limits_{#1}}

\newcommand{\maxi}[1]{\mathop{\mathrm{maximize}}\limits_{#1}}

\newcommand{\st}{\mathrm{subject}\quad\mathrm{to}}
\newcommand{\ith}[1]{$#1$th}

\newcommand{\toinf}[1]{#1 \to\infty}

\newcommand{\delfunc}[1]{\delta\left(#1\right)}

\newcommand{\vecele}[2]{\left[#1\right]_{#2}}

\newcommand{\setdif}[2]{#1\backslash #2}

\newcommand{\thetabs}[2]{{\dnnot{\theta}{bs}}}

\newcommand{\parenth}[1]{\left(#1\right)}

\newcommand{\equaa}{\mathop{=}^{(\mathrm{a})}}
\newcommand{\equab}{\mathop{=}^{(\mathrm{b})}}
\newcommand{\equac}{\mathop{=}^{(\mathrm{c})}}
\newcommand{\equad}{\mathop{=}^{(\mathrm{d})}}

\newcommand{\cB}{\mathcal{B}}
\newcommand{\cC}{\mathcal{C}}

\newcommand{\cS}{\mathcal{S}}

\newcommand{\cU}{\mathcal{U}}

\newcommand{\ba}{\mathbf{a}}

\newcommand{\bv}{\mathbf{v}}

\newcommand{\bx}{\mathbf{x}}
\newcommand{\by}{\mathbf{y}}
\newcommand{\bz}{\mathbf{z}}

\newcommand{\bA}{\mathbf{A}}
\newcommand{\bB}{\mathbf{B}}

\newcommand{\bG}{\mathbf{G}}

\newcommand{\bI}{\mathbf{I}}

\newcommand{\bQ}{\mathbf{Q}}

\newcommand{\bU}{\mathbf{U}}
\newcommand{\bV}{\mathbf{V}}

\newcommand{\bbC}{\mathbb{C}}
\newcommand{\bbR}{\mathbb{R}}

\newcommand{\ttS}{\mathtt{S}}

\newcommand{\barx}{\overline{x}}
\newcommand{\bary}{\overline{y}}

\newcommand{\barbG}{\overline{\bG}}

\newcommand{\barbx}{\overline{\bx}}
\newcommand{\barby}{\overline{\by}}
\newcommand{\barbz}{\overline{\bz}}

\newcommand{\barbQ}{\overline{\bQ}}
\newcommand{\barbU}{\overline{\bU}}

\newcommand{\bzero}{\mathbf{0}}

\newcommand{\blambda}{{\boldsymbol\lambda}}

\newcommand{\bOmega}{{\boldsymbol\Omega}}

\newcommand{\sintheta}{\sin\left(\theta\right)}

\newcommand{\sinphi}{\sin\left(\phi\right)}

\newcommand{\expx}[1]{\exp\left\{#1\right\}}

\newcommand{\upnot}[2]{#1^{\mathrm{#2}}}
\newcommand{\dnnot}[2]{#1_{\mathrm{#2}}}

\newcommand{\upul}[1]{#1^{\mathrm{ul}}}
\newcommand{\updl}[1]{#1^{\mathrm{dl}}}

\newcommand{\zetaul}{{\zeta}_{\mathrm{ul}}}
\newcommand{\zetadl}{{\zeta}_{\mathrm{dl}}}

\newcommand{\bGdl}{{\bG}^{\mathrm{dl}}}

\newcommand{\barbxul}{{\barbx}^{\mathrm{ul}}}
\newcommand{\barbxdl}{{\barbx}^{\mathrm{dl}}}
\newcommand{\barbzul}{{\barbz}^{\mathrm{ul}}}
\newcommand{\barbzdl}{{\barbz}^{\mathrm{dl}}}

\newcommand{\barbGul}{{\barbG}^{\mathrm{ul}}}
\newcommand{\barbGdl}{{\barbG}^{\mathrm{dl}}}

\newcommand{\barbQdl}{{\barbQ}^{\mathrm{dl}}}

\newcommand{\barjmath}{\bar{\jmath}}

\newcommand{\ntb}{\notag\\}

\allowdisplaybreaks

\begin{document}

\title{\LARGE BDMA for Millimeter-Wave/Terahertz Massive MIMO Transmission with Per-Beam Synchronization}

\author{
Li~You, Xiqi~Gao, Geoffrey~Ye~Li, Xiang-Gen~Xia, and~Ni~Ma%
\thanks{
Part of this work has been submitted to IEEE ICC'17.
}
\thanks{
L. You and X. Q. Gao are with the National Mobile Communications Research Laboratory, Southeast University, Nanjing 210096, China (e-mail: lyou@seu.edu.cn; xqgao@seu.edu.cn).
}
\thanks{
G. Y. Li is with the School of Electrical and Computer Engineering, Georgia Institute of Technology, Atlanta, GA 30332 USA (e-mail: liye@ece.gatech.edu).
}
\thanks{
X.-G. Xia is with the Department of Electrical and Computer Engineering, University of Delaware, Newark, DE 19716 USA (e-mail: xxia@ee.udel.edu).
}
\thanks{
N. Ma is with the Huawei Technologies Co., Ltd., Shenzhen 518129, China (e-mail: ni.ma@huawei.com).
}
}

\maketitle

\begin{abstract}
We propose beam division multiple access (BDMA) with per-beam synchronization (PBS) in time and frequency for wideband massive multiple-input multiple-output (MIMO) transmission over millimeter-wave (mmW)/Terahertz (THz) bands.
We first introduce a physically motivated beam domain channel model for massive MIMO and demonstrate that the envelopes of the beam domain channel elements tend to be independent of time and frequency when both the numbers of antennas at base station and user terminals (UTs) tend to infinity.
Motivated by the derived beam domain channel properties, we then propose PBS for mmW/THz massive MIMO. We show that both the effective delay and Doppler frequency spreads of wideband massive MIMO channels with PBS are reduced by a factor of the number of UT antennas compared with the conventional synchronization approaches.
Subsequently, we apply PBS to BDMA, investigate beam scheduling to maximize the achievable ergodic rates for both uplink and downlink BDMA, and develop a greedy beam scheduling algorithm.
Simulation results verify the effectiveness of BDMA with PBS for mmW/THz wideband massive MIMO systems in typical mobility scenarios.
\end{abstract}

\newpage

\section{Introduction}\label{sec:bm_intro}

With severe spectrum shortage in the currently deployed cellular bands (sub-6 GHz) and the explosive wireless traffic demand, there is a growing consensus on utilizing higher frequency bands, e.g., the millimeter-wave (mmW) band and the Terahertz (THz) band, for future wireless communication systems \cite{Rangan14Millimeter,Heath16overview,Akyildiz14TeraNets,Han16Exploiting,Lin16Energy}.
Massive multiple-input multiple-output (MIMO) transmission deploys large numbers of antennas at the base stations (BSs) to simultaneously serve multiple user terminals (UTs) and can significantly improve the system spectrum efficiency \cite{Marzetta10Noncooperative,Lu14overview}.
Combination of massive MIMO with mmW/THz technologies is appealing from a practical point of view. Orders-of-magnitude smaller wavelength in mmW/THz bands enables a larger number of antennas to be deployed at both UTs and BSs.
Even for a high propagation path loss at mmW/THz channels, the achievable high beamforming gains with massive MIMO can help to compensate for it.
Therefore, massive MIMO transmission over mmW/THz bands, which will be referred to as mmW/THz massive MIMO, is envisioned as a promising solution for wireless communications in the future \cite{Swindlehurst14Millimeter,Akyildiz14TeraNets}.

Utilizing mmW/THz frequencies for cellular wireless has received intense research interest recently. One challenge in realizing cellular wireless over mmW/THz channels is to deal with the mobility issue \cite{Roh14Millimeter,Andrews16Modeling}. For the same mobile speed, the Doppler spread of mmW/THz channels is orders-of-magnitude larger than that of classical wireless channels while the delay spread does not change significantly over different frequencies, which may lead to system implementation bottleneck. Consider wideband mmW/THz transmission employing orthogonal frequency division multiplexing (OFDM) modulation for example. With perfect time and frequency synchronization in the space domain, the length of the cyclic prefix (CP) is usually set to be slightly larger than the delay span to mitigate channel dispersion in time while the length of the OFDM symbol is usually set to be inversely proportional to the Doppler spread to mitigate channel dispersion in frequency \cite{Hwang09OFDM,Dahlman14LTE}.\footnote{While it is possible to relax these requirements and mitigate the negative effects by advanced algorithms, such approaches are not considered here due to the relatively high implementation complexity.} As a result, the overhead of the CP will be much larger to deal with the same delay spread and it might be difficult to design proper OFDM parameters.

There exist some works related to the above issue. For example, beam-based Doppler frequency compensation has been suggested in \cite{Chizhik04Slowing} and \cite{Va16impact} for narrowband MIMO channels. In addition, reduced delay spread with narrow directional beams has been observed in recent mmW channel measurement results \cite{Rappaport15Wideband}.

In this paper, we exploit massive MIMO to address the above issue.
Specifically, we propose per-beam synchronization (PBS) for mmW/THz massive MIMO-OFDM transmission and apply it to the recently proposed beam division multiple access (BDMA) \cite{Sun15Beam}. The major contributions of this paper are summarized as follows, part of which has been submitted to a conference \cite{You17Millimeter}:
\begin{itemize}
\item We introduce a physically motivated beam domain channel model for massive MIMO. We show that when both the numbers of antennas at BS and UTs are sufficiently large, the beam domain channel elements tend to be statistically uncorrelated and the respective variances depend on the channel power angle spectra (PAS), while the envelopes of the beam domain channel elements tend to be independent of time and frequency.
\item We propose PBS in time and frequency for mmW/THz massive MIMO. Both delay and Doppler frequency spreads of the wideband MIMO channels with PBS are shown to be approximately reduced by a factor of the number of UT antennas compared with the conventional synchronization approaches \cite{Hwang09OFDM}. Note that the proposed PBS can also be applied to massive MIMO transmission over other frequency bands as long as the numbers of UT antennas are sufficiently large.
\item We apply PBS to BDMA in which multiple access is achieved by providing each UT with a mutually non-overlapping subset of BS beams \cite{Sun15Beam}. We investigate beam scheduling to maximize achievable ergodic rates for both uplink (UL) and downlink (DL) BDMA, and develop a greedy beam scheduling algorithm based on the average squared beam domain channel norm.
\end{itemize}

The rest of this paper is organized as follows. In \secref{sec:bm_channel}, we investigate the beam domain channel model. In \secref{sec:bm_sync}, we propose PBS for mmW/THz massive MIMO. In \secref{sec:bdma_pbs_syn}, we apply PBS to BDMA. Simulation results are presented in \secref{sec:pbs_sim} and the paper is concluded in \secref{sec:pbs_conc}.

Some of the notations used in this paper are listed as follows:
\begin{itemize}
\item $\barjmath=\sqrt{-1}$. $\delta(\cdot)$ denotes the delta function.
\item $\bbC^{M\times N}$ ($\bbR^{M\times N}$) denotes the $M\times N$ dimensional complex (real) vector space.
\item Upper and lower case boldface letters denote matrices and column vectors, respectively.
\item ${\bI}_{N}$ denotes the $N\times N$ dimensional identity matrix, and the subscript is sometimes omitted for brevity. $\bzero$ denotes the all-zero vector (matrix).
\item $(\cdot)^{H}$, $(\cdot)^{T}$, and $(\cdot)^{*}$ denote conjugate transpose, transpose, and conjugate operations, respectively.
\item $\diag{\bx}$ denotes the diagonal matrix with $\bx$ along its main diagonal. $\tr{\cdot}$ denotes the matrix trace operation.
\item $\vecele{\ba}{i}$ and $\vecele{\bA}{i,j}$ denote the \ith{i} element of $\ba$, and the $(i,j)$th element of $\bA$, respectively, where the element indices start from $0$. $\vecele{\bA}{\cB,:}$, $\vecele{\bA}{:,\cC}$, and $\vecele{\bA}{\cB,\cC}$ denote the submatrices of $\bA$ consisting of rows specified in $\cB$ and (or) columns specified in $\cC$.
\item $\expect{\cdot}$ denotes the expectation operation. $\GCN{\ba}{\bB}$ denotes the circular symmetric complex Gaussian distribution with mean $\ba$ and covariance $\bB$.
\item $\backslash$ denotes the set subtraction operation. $\abs{\cB}$ denotes the cardinality of set $\cB$.
\item $\triangleq$ denotes ``be defined as''. $\sim$ denotes ``be distributed as''.
\end{itemize}

\section{Beam Domain Channel Model}\label{sec:bm_channel}

In this section, we will first introduce a physically motivated beam domain channel model for mmW/THz massive MIMO and then investigate its properties.

Consider a single-cell massive MIMO system, where the BS with $M$ antennas simultaneously serves $U$ UTs, each with $K$ antennas. The UT set is denoted as $\cU=\left\{\zeroldots{U-1}\right\}$ where $u\in\cU$ denotes the UT index.
The small wavelength in mmW/THz bands makes it possible to pack a large number of antennas at the UTs in addition to the BS. We focus on the case where both the numbers of antennas at the BS and the UTs are sufficiently large, which is different from the massive MIMO communications over lower frequency bands \cite{Marzetta10Noncooperative}.

\subsection{DL Channel Model}

We assume that both the BS and the UTs are equipped with uniform linear arrays (ULAs) with one-half wavelength antenna spacing. The array response vectors corresponding to the angles of departure/arrival (AoD/AoA) with respect to the perpendicular to the BS and the UT arrays are given by \cite{Tse05Fundamentals}
\begin{align}
\dnnot{\bv}{bs}\left(\theta\right)&=\left[1\quad\expx{-\barjmath\pi\sintheta}\quad\ldots\quad\expx{-\barjmath\pi(M-1)\sintheta}\right]^{T}\in\bbC^{M\times 1},\label{eq:ula steer vec_bs}\\
\dnnot{\bv}{ut}\left(\phi\right)&=\left[1\quad\expx{-\barjmath\pi\sinphi}\quad\ldots\quad\expx{-\barjmath\pi(K-1)\sinphi}\right]^{T}\in\bbC^{K\times 1},\label{eq:ula steer vec_ut}
\end{align}
respectively.
As indicated in \cite{Chizhik04Slowing,You15Pilot}, the front-to-back ambiguity of the linear array can usually be mitigated via proper configurations. Therefore, we assume that the angles $\theta$ and $\phi$ lie in interval $[-\pi/2,\pi/2]$ without loss of generality.

We assume that the channels between the BS and different UTs are uncorrelated and focus on the DL channel between the BS and UT $u$.
For the ray-tracing based wireless channel model \cite{Tse05Fundamentals}, the received signal is constituted of a sum of the multiple transmitted signal copies, experiencing different attenuations, AoAs, AoDs, Doppler shifts, and delays.

The channel delay and Doppler shift properties are usually related to its AoA-AoD properties \cite{Sayeed02essential,Rappaport12angle,Gustafson14On}.
We first consider the relationship between the Doppler shift and the AoA-AoD pair. Assume that the scatterers are stationary and the channel temporal fluctuation is mainly due to the motion of the UT. Also assume that UT $u$ moves along a straight line at a constant velocity $v_{u}$ and the motion direction is parallel to the ULA of UT $u$. Then following the Clarke-Jakes model \cite{Patzold12Mobile}, the channel path with AoA $\phi$ will experience a Doppler shift $\nu_u\left(\phi\right)$ as
\begin{align}\label{eq:dop_jakes}
  \nu_u\left(\phi\right)=\nu_{u}\sinphi,
\end{align}
where $\nu_u\triangleq \dnnot{f}{c}v_u/c$ is the maximum Doppler shift of UT $u$, $\dnnot{f}{c}$ is the carrier frequency,\footnote{Note that the Doppler shift, $\nu_u\left(\phi\right)$, is usually assumed to be constant over the frequency band of interest in practical wireless systems, although rigourously speaking it is a function of the actual operating frequency \cite{Gallager08Principles}.} and $c$ is the light speed.

Consider the relationship between the propagation delay and the AoA-AoD pair. Due to the channel sparsity \cite{Rangan14Millimeter} and the relatively large transmission bandwidth over mmW/THz bands, the probability that two resolvable propagation paths have the same AoA-AoD pair but different path delays can be almost neglected \cite{Zeng16Millimeter}.
Therefore, we assume that there are no two paths with the same AoA-AoD pair but different path delays, and the path delay of the channel with AoA-AoD pair $\left(\phi,\theta\right)$ is defined as $\tau_u\left(\phi,\theta\right)$.

With the above modeling of the channel delay and Doppler shift, the corresponding complex baseband DL space domain channel frequency response, $\bGdl_u\parenth{t,f}$, at time $t$ and frequency $f$ can be represented as (see, e.g., \cite{Tse05Fundamentals,Clerckx13MIMO,Auer12MIMO,You16Channel})
\begin{align}\label{eq:con_wb_cha_mod_rg_del}
\bGdl_{u}\parenth{t,f}&=\int\limits_{-\frac{\pi}{2}}^{\frac{\pi}{2}}\int\limits_{-\frac{\pi}{2}}^{\frac{\pi}{2}}\!
\sqrt{\ttS_{u}\left(\phi,\theta\right)}
\cdot\expx{\barjmath\zetadl\parenth{\phi,\theta}}
\cdot\dnnot{\bv}{ut}\left(\phi\right)\dnnot{\bv}{bs}^{T}\left(\theta\right)
\ntb
 &\qquad
\cdot\expx{\barjmath2\pi\left[t\nu_u\left(\phi\right)-f\tau_u\left(\phi,\theta\right)\right]}
  \intdx{\phi}\intdx{\theta}\in\bbC^{K\times M},
\end{align}
where $\ttS_{u}\left(\phi,\theta\right)$ is the average power of the path associated with AoD-AoA pair $\parenth{\phi,\theta}$ given by the PAS of UT $u$, and $\zetadl\parenth{\phi,\theta}$ is a random phase that is uniformly distributed over $\left[0,2\pi\right)$ and independent of $\zetadl\parenth{\phi',\theta'}$ for $\phi\neq\phi'$ or $\theta\neq\theta'$.
Note that the channel model in \eqref{eq:con_wb_cha_mod_rg_del} has been widely adopted and verified in recent mmW/THz works \cite{Lin16Energy,Akdeniz14Millimeter}.
Also, the above channel model applies over time intervals where the relative positions of the UTs do not change significantly and the physical channel parameters, $\nu_u\left(\phi\right)$, $\tau_u\left(\phi,\theta\right)$, and $\ttS_{u}\left(\phi,\theta\right)$, can be assumed to be time-invariant. When the positions of the UTs change significantly, these parameters should be updated accordingly \cite{Clerckx13MIMO}.

Following the MIMO channel modeling approach in \cite{Sayeed02Deconstructing,You15Pilot,You16Channel}, we define
\begin{align}\label{eq:angchaaoaaod1}
\barbGdl_{u}\parenth{t,f}&\triangleq\bV_{K}^{H}\bGdl_{u}\parenth{t,f}\bV_{M}^{*}\in\bbC^{K\times M},
\end{align}
where $\bV_{K}\in\bbC^{K\times K}$ with $\vecele{\bV_{K}}{i,j}\triangleq1/\sqrt{K}\cdot\expx{-\barjmath2\pi i\left(j-K/2\right)/K}$ is the unitary discrete Fourier transform (DFT) matrix (with matrix elementary operations). Both transformation matrices, $\bV_{K}$ and $\bV_{M}$, in \eqref{eq:angchaaoaaod1} can be interpreted as DFT beamforming operations performed at the BS and the UTs, respectively. Thus, we refer to $\barbGdl_{u}\parenth{t,f}$ as the DL beam domain channel frequency response matrix between the BS and UT $u$ at time $t$ and frequency $f$.

\subsection{Asymptotic DL Channel Properties}

From \eqref{eq:angchaaoaaod1}, the elements of the DL beam domain channel between the BS and UT $u$ can be written as
\begin{align}\label{eq:beam_cha_ele}
  &\vecele{\barbGdl_{u}\parenth{t,f}}{k,m}=\vecele{\bV_{K}}{:,k}^{H}\bGdl_{u}\parenth{t,f}\vecele{\bV_{M}}{:,m}^{*}
  \ntb
  &\qquad\equaa
  \int\limits_{-\frac{\pi}{2}}^{\frac{\pi}{2}}\int\limits_{-\frac{\pi}{2}}^{\frac{\pi}{2}}\!
  \sqrt{\ttS_{u}\left(\phi,\theta\right)}
\cdot\expx{\barjmath\zetadl\parenth{\phi,\theta}}
  \cdot\expx{\barjmath2\pi\left[t\nu_u\left(\phi\right)-f\tau_u\left(\phi,\theta\right)\right]}
  \ntb
  &\qquad\qquad
  \cdot
  \vecele{\bV_{K}}{:,k}^{H}\dnnot{\bv}{ut}\left(\phi\right)\dnnot{\bv}{bs}^{T}\left(\theta\right)
  \vecele{\bV_{M}}{:,m}^{*}
  \intdx{\phi}\intdx{\theta}
  \ntb
  &\qquad\equab
  \int\limits_{-\frac{\pi}{2}}^{\frac{\pi}{2}}\int\limits_{-\frac{\pi}{2}}^{\frac{\pi}{2}}\!
\sqrt{\ttS_{u}\left(\phi,\theta\right)}
\cdot\expx{\barjmath\zetadl\parenth{\phi,\theta}}
  \cdot\expx{\barjmath2\pi\left[t\nu_u\left(\phi\right)-f\tau_u\left(\phi,\theta\right)\right]}
  \ntb
  &\qquad\qquad
  \cdot
  \oneon{\sqrt{K}}\sum_{a=0}^{K-1}\expx{\barjmath\pi a\left[\left(\frac{2k}{K}-1\right)-\sinphi\right]}
   \ntb
  &\qquad\qquad
  \cdot\oneon{\sqrt{M}}\sum_{b=0}^{M-1}\expx{\barjmath\pi b\left[\left(\frac{2m}{M}-1\right)-\sintheta\right]}
  \intdx{\phi}\intdx{\theta}
  \ntb
  &\qquad\equac
  \int\limits_{-\frac{\pi}{2}}^{\frac{\pi}{2}}\int\limits_{-\frac{\pi}{2}}^{\frac{\pi}{2}}\!
\sqrt{\ttS_{u}\left(\phi,\theta\right)}
\cdot\expx{\barjmath\zetadl\parenth{\phi,\theta}}
  \cdot\expx{\barjmath2\pi\left[t\nu_u\left(\phi\right)-f\tau_u\left(\phi,\theta\right)\right]}
  \ntb
  &\qquad\qquad
  \cdot
  q_K\left(\left(\frac{2k}{K}-1\right)-\sinphi\right)\cdot
  q_M\left(\left(\frac{2m}{M}-1\right)-\sintheta\right)
  \intdx{\phi}\intdx{\theta},
\end{align}
where (a) follows from \eqref{eq:con_wb_cha_mod_rg_del}, (b) follows from \eqref{eq:ula steer vec_bs} and \eqref{eq:ula steer vec_ut}, and (c) follows from the definition
\begin{align}\label{eq:beam_samp}
  q_K\left(x\right)\triangleq\oneon{\sqrt{K}}\sum_{k=0}^{K-1}\expx{\barjmath\pi kx}=\expx{\barjmath\frac{\pi}{2}\left(K-1\right)x}\frac{\sin\left(\frac{\pi}{2}Kx\right)}{\sqrt{K}\sin\left(\frac{\pi}{2}x\right)}.
\end{align}
Note that $q_K\left(x\right)$ tends to the delta function when $K$ tends to infinity, then an asymptotic property of the beam domain channel frequency response matrix can be stated in the following proposition.

\begin{prop}\label{prop:adcmsts_mimo}
Define $\upnot{\barbG_{u}}{dl,asy}\parenth{t,f}\in\bbC^{K\times M}$ as
\begin{align}\label{eq:exbhk_mimo_aymp}
  \vecele{\upnot{\barbG_{u}}{dl,asy}\parenth{t,f}}{k,m}&\triangleq
  \sqrt{\ttS_{u}\left(\phi_k,\theta_m\right)}
  \cdot\expx{\barjmath\zetadl\parenth{\phi_k,\theta_m}}
  \cdot\expx{\barjmath2\pi\left[t\nu_{u}\left(\phi_k\right)-f\tau_u\left(\phi_k,\theta_m\right)\right]},
\end{align}
where
\begin{align}\label{eq:vir_ang}
  \phi_{k}\triangleq\arcsin\left(\frac{2k}{K}-1\right),
  \ \
  \theta_{m}\triangleq\arcsin\left(\frac{2m}{M}-1\right).
\end{align}
Then $\barbGdl_{u}\parenth{t,f}\to\upnot{\barbG_{u}}{dl,asy}\parenth{t,f}$ in the sense that, for fixed non-negative integers $k$ and $m$,
\begin{align}\label{eq:exbhk_mimo}
  \lim_{\substack{\toinf{K,M}}}\vecele{
  \barbGdl_{u}\parenth{t,f}-\upnot{\barbG_{u}}{dl,asy}\parenth{t,f}}{k,m}&=0.
\end{align}
\end{prop}

\propref{prop:adcmsts_mimo} shows that the beam domain channel elements asymptotically tend to exhibit the structures as in \eqref{eq:exbhk_mimo_aymp} when the numbers of antennas tend to infinity. With \propref{prop:adcmsts_mimo}, we proceed to investigate the beam domain channel properties in the large array regime.
Directly from \eqref{eq:exbhk_mimo_aymp} and the assumption that the random phases $\zetadl\parenth{\phi_k,\theta_m}$ and $\zetadl\parenth{\phi_{k'},\theta_{m'}}$ are independent for $k\neq k'$ or $m\neq m'$, the following proposition on the statistics of the beam domain channels can be obtained.

\begin{prop}\label{prop:beam_cha_stat}
Define $\upnot{\bOmega_{u}}{asy}\in\bbR^{K\times M}$ as
\begin{align}\label{eq:cov_mimo_aymp}
  \vecele{\upnot{\bOmega_{u}}{asy}}{k,m}\triangleq
  \ttS_u\parenth{\phi_k,\theta_m}.
\end{align}
Then, for every $t$ and every $f$, when the numbers of antennas, $M$ and $K$, both tend to infinity, the beam domain channel elements satisfy
\begin{align}\label{eq:beam_cha_stat}
  \expect{\vecele{\barbGdl_u\parenth{t,f}}{k,m}\vecele{\barbGdl_u\parenth{t,f}}{k',m'}^{*}}
  &\to \vecele{\upnot{\bOmega_{u}}{asy}}{k,m}\cdot\delfunc{k-k'}\delfunc{m-m'}.
\end{align}
\end{prop}

\propref{prop:beam_cha_stat} shows that different beam domain channel elements are asymptotically uncorrelated. In addition, the variances of the beam domain channel elements are independent of the frequency, $f$, and are related to the corresponding channel PAS, which lends the beam domain channel matrix defined in \eqref{eq:angchaaoaaod1} its physical interpretation. Specifically, different beam domain channel elements correspond to the channel gains of different transmit-receive beam directions, which can be resolved in mmW/THz massive MIMO with sufficiently large antenna array apertures at both the BS and the UT sides.
Note that mmW/THz channels usually exhibit an approximately sparse nature compared with channels over regular bands \cite{Andrews16Modeling} and most of the elements in $\upnot{\bOmega_{u}}{asy}$ are approximately zero, which can be exploited to facilitate wireless transmission design.

Note that \propref{prop:beam_cha_stat} coincides with many of the existing results. For example, the result that the space domain channel statistics are independent of frequency has been shown in \cite{Liu03Capacity} while our result is established in the beam domain. In addition, for the case with single-antenna UTs, the result in \eqref{eq:beam_cha_stat} has been shown to be accurate enough for a practical number of antennas at the BS \cite{Wen15Channel,Yin13coordinated,Adhikary13Joint,You15Pilot,You16Channel} while our result corresponds to the case where the UTs are also equipped with a large number of antennas, which is of practical interest for massive MIMO communications over mmW/THz bands.

From \eqref{eq:exbhk_mimo_aymp}, the dispersion property of the beam domain channels can be obtained as follows.

\begin{prop}\label{prop:beam_cha_flat}
Define $\upnot{\barbG}{asy,env}_u\in\bbR^{K\times M}$ as
\begin{align}\label{eq:beam_env}
  \vecele{\upnot{\barbG}{asy,env}_u}{k,m}
  \triangleq\sqrt{\ttS_{u}\left(\phi_k,\theta_m\right)}.
\end{align}
Then when the numbers of antennas, $M$ and $K$, both tend to infinity, the envelopes of the beam domain channel elements tend to be independent of the time, $t$, and the frequency, $f$, in the sense that, for every $t$ and every $f$ and for fixed non-negative integers $k$ and $m$,
\begin{align}\label{eq:beam_flat}
  \abs{\vecele{\barbGdl_u\parenth{t,f}}{k,m}}
  &\to\vecele{\upnot{\barbG}{asy,env}_u}{k,m}.
\end{align}
\end{prop}

From \propref{prop:beam_cha_flat}, in the asymptotically large array regime, the fading of each of the beam domain channel elements tends to disappear when both the numbers of antennas at the BS and the UTs tend to infinity. The physical interpretation of \propref{prop:beam_cha_flat} is intuitive. Specifically, beamforming can effectively divide the channels in the angle domain and the partition resolution can be sufficiently high with sufficiently large numbers of antennas at both the BS and the UTs.
Asymptotically, each propagation path can be resolved and the beam domain channel element corresponds to the gain of a specific propagation path along a fixed AoA-AoD pair. Thus, the beam domain channel envelopes tend to remain a constant in both the time and the frequency domains. Note that a narrowband case of \propref{prop:beam_cha_flat} has been obtained in \cite{Chizhik04Slowing}. In addition, reduced delay spread with beamforming has been observed in recent mmW channel measurement results \cite{Rappaport15Wideband}.
For wideband mmW/THz massive MIMO channels, we take into account of both delay and Doppler spreading.

\subsection{DL Channel Approximation}

We have derived several asymptotic properties of the beam domain channels above. Before proceeding, we investigate the case with finite (but large) numbers of antennas.
Note that function $q_K(x)$ defined in \eqref{eq:beam_samp} has a sharper peak around $x=0$ with a larger $K$ \cite{Sayeed02Deconstructing,Tse05Fundamentals}. Thus, for sufficiently large $K$ and $M$, the beam domain channel elements in \eqref{eq:beam_cha_ele} can be well approximated by
\begin{align}\label{eq:beam_cha_ele_appr}
  \vecele{\barbGdl_{u}\parenth{t,f}}{k,m}
  &\simeq
  \int\limits_{\theta_m}^{\theta_{m+1}}\int\limits_{\phi_k}^{\phi_{k+1}}\!
\sqrt{\ttS_{u}\left(\phi,\theta\right)}
\cdot\expx{\barjmath\zetadl\parenth{\phi,\theta}}
\ntb
  &\qquad\cdot\expx{\barjmath2\pi\left[t\nu_u\left(\phi\right)-f\tau_u\left(\phi,\theta\right)\right]}
  \intdx{\phi}\intdx{\theta}.
\end{align}

The approximation in \eqref{eq:beam_cha_ele_appr} coincides with the physical intuition of the beam domain channels. Specifically, with a larger number of antennas, the antenna array has the ability of forming narrower beams. For a given transmit-receive beam pair, the transmitted signals will be focused on the corresponding AoA-AoD pairs meanwhile the signal leakage can be almost neglected. It is also worth noting that most of the beam domain channel properties in the asymptotic regime are well reflected in
the approximation model given in \eqref{eq:beam_cha_ele_appr}. For example, the delay and Doppler spreads of the whole beam domain channel matrices are maintained. Meanwhile, the delay and Doppler spreads of a specific beam domain channel element tend to disappear, which coincides with \propref{prop:beam_cha_flat}.

The approximated DL beam domain channel elements in \eqref{eq:beam_cha_ele_appr} are uncorrelated in the sense that
\begin{align}\label{eq:beam_cha_pow}
\expect{\vecele{\barbGdl_{u}\parenth{t,f}}{k,m}\vecele{\barbGdl_{u}\parenth{t,f}}{k',m'}^{*}}
=\underbrace{\int\limits_{\theta_m}^{\theta_{m+1}}\int\limits_{\phi_k}^{\phi_{k+1}}\!\ttS_{u}\left(\phi,\theta\right)
  \intdx{\phi}\intdx{\theta}}_{\triangleq\vecele{\bOmega_u}{k,m}}\cdot\delfunc{k-k'}\delfunc{m-m'},
\end{align}
where $\bOmega_u\in\bbR^{K\times M}$ is referred to as the beam domain channel power matrix, which coincides with \propref{prop:beam_cha_stat}. In the rest of the paper, we will thus exclusively use the simplified channel model in \eqref{eq:beam_cha_ele_appr}.

\subsection{UL Channel Model}

In the above subsections, we investigate the DL beam domain channel properties. Hereafter we briefly discuss the UL case.

For time-division duplex systems, the UL channel response is the transpose of the DL channel response at the same time and frequency. Thus, similar results as presented in the above subsections can be readily obtained.

For frequency-division duplex systems where the relative carrier frequency difference is small, the physical channel parameters, $\ttS_{u}\left(\phi,\theta\right)$, $\nu_{u}\parenth{\phi}$, and $\tau_{u}\parenth{\phi,\theta}$, as well as the array responses are almost identical for both the UL and DL \cite{Xu04generalized,Ugurlu16multipath,Barriac04Space}.
Thus, the major difference between the UL and DL channels lies in the random phase term, and the UL beam domain channel frequency response matrix between UT $u$ and the BS at time $t$ and frequency $f$, $\barbGul_{u}\parenth{t,f}\in\bbC^{M\times K}$, can be modeled as
\begin{align}\label{eq:con_wb_cha_mod_rg_del_ul}
\barbGul_{u}\parenth{t,f}
&\triangleq\int\limits_{-\frac{\pi}{2}}^{\frac{\pi}{2}}\int\limits_{-\frac{\pi}{2}}^{\frac{\pi}{2}}\!
\sqrt{\ttS_{u}\left(\phi,\theta\right)}
\cdot\expx{\barjmath\zetaul\parenth{\phi,\theta}}
\cdot
\bV_{M}^{H}\dnnot{\bv}{bs}\left(\theta\right)\dnnot{\bv}{ut}^{T}\left(\phi\right)\bV_{K}^{*}
\ntb
 &\qquad
\cdot\expx{\barjmath2\pi\left[t\nu_u\left(\phi\right)-f\tau_u\left(\phi,\theta\right)\right]}
  \intdx{\phi}\intdx{\theta}\in\bbC^{M\times K},
\end{align}
where $\zetaul\parenth{\phi,\theta}$ is a random phase in the UL
that is uncorrelated with the DL random phase $\zetadl\parenth{\phi,\theta}$, uniformly distributed over $\left[0,2\pi\right)$, and
independent of $\zetaul\parenth{\phi',\theta'}$ for $\phi\neq\phi'$ or $\theta\neq\theta'$.

Similarly as \eqref{eq:beam_cha_ele_appr}, for sufficiently large $K$ and $M$, $\barbGul_{u}\parenth{t,f}$ can be well approximated by
\begin{align}\label{eq:beam_cha_ele_appr_ul}
  \vecele{\barbGul_{u}\parenth{t,f}}{m,k}
  &\simeq
  \int\limits_{\theta_m}^{\theta_{m+1}}\int\limits_{\phi_k}^{\phi_{k+1}}\!
\sqrt{\ttS_{u}\left(\phi,\theta\right)}
\cdot\expx{\barjmath\zetaul\parenth{\phi,\theta}}
\ntb
  &\qquad\cdot\expx{\barjmath2\pi\left[t\nu_u\left(\phi\right)-f\tau_u\left(\phi,\theta\right)\right]}
  \intdx{\phi}\intdx{\theta}.
\end{align}
From \eqref{eq:beam_cha_ele_appr_ul} and the definitions of $\phi_i$ and $\theta_j$ in \eqref{eq:vir_ang}, the delay and Doppler spreads of the UL beam domain channels $\vecele{\barbGul_{u}\parenth{t,f}}{m,k}$ tend to decrease with increasing $K$ and $M$.
In addition, from \eqref{eq:beam_cha_pow} and \eqref{eq:beam_cha_ele_appr_ul}, the UL beam domain channel elements are uncorrelated in the sense that
\begin{align}\label{eq:beam_cha_stat_ul}
\expect{\vecele{\barbGul_u\parenth{t,f}}{m,k}\vecele{\barbGul_u\parenth{t,f}}{m',k'}^{*}}
  &=\vecele{\bOmega_{u}}{k,m}\cdot\delfunc{k-k'}\delfunc{m-m'},
\end{align}
which reveals the reciprocity between the UL and DL beam domain channel statistics.

Before concluding this section, we define the (approximated) DL and UL beam domain channel impulse response matrices, $\barbGdl_{u}\parenth{t,\tau}\in\bbC^{K\times M}$ and $\barbGul_{u}\parenth{t,\tau}\in\bbC^{M\times K}$, as the inverse Fourier transforms of $\barbGdl_{u}\parenth{t,f}$ and $\barbGul_{u}\parenth{t,f}$ given by
\begin{align}
  \vecele{\barbGdl_{u}\parenth{t,\tau}}{k,m}
  &=
  \int\limits_{\theta_m}^{\theta_{m+1}}\int\limits_{\phi_k}^{\phi_{k+1}}\!
\sqrt{\ttS_{u}\left(\phi,\theta\right)}
\cdot\expx{\barjmath\zetadl\parenth{\phi,\theta}}
\ntb
  &\qquad\cdot\expx{\barjmath2\pi t\nu_u\left(\phi\right)}
\cdot\delfunc{\tau-\tau_u\left(\phi,\theta\right)}
  \intdx{\phi}\intdx{\theta},\label{eq:beam_cha_ele_appr_del}\\
\vecele{\barbGul_{u}\parenth{t,\tau}}{m,k}
  &=
  \int\limits_{\theta_m}^{\theta_{m+1}}\int\limits_{\phi_k}^{\phi_{k+1}}\!
\sqrt{\ttS_{u}\left(\phi,\theta\right)}
\cdot\expx{\barjmath\zetaul\parenth{\phi,\theta}}
\ntb
  &\qquad\cdot\expx{\barjmath2\pi t\nu_u\left(\phi\right)}
\cdot\delfunc{\tau-\tau_u\left(\phi,\theta\right)}
  \intdx{\phi}\intdx{\theta},\label{eq:beam_cha_ele_appr_del_ul}
\end{align}
respectively, which will be adopted to simplify the analyses in the following section.

\section{PBS in Time and Frequency}\label{sec:bm_sync}

In the above section, we have investigated the beam domain channel model for mmW/THz massive MIMO. Based on the obtained beam domain channel properties, in this section we propose PBS in time and frequency for mmW/THz massive MIMO communications to reduce the effective MIMO channel dispersion in time and frequency. We first investigate DL synchronization, and then briefly address the UL case.

\subsection{DL Transmission Model}

We consider an mmW/THz wideband massive MIMO system employing OFDM modulation with the number of subcarriers, $\dnnot{N}{us}$, and the CP, $\dnnot{N}{cp}$ samples.
Then the OFDM symbol length and the CP length are $\dnnot{T}{us}=\dnnot{N}{us}\dnnot{T}{s}$ and $\dnnot{T}{cp}=\dnnot{N}{cp}\dnnot{T}{s}$, respectively, where $\dnnot{T}{s}$ is the system sampling interval.

Let $\left\{\upnot{\barbx}{dl}_{n}\right\}_{n=0}^{\dnnot{N}{us}-1}$ be the complex-valued symbols to be transmitted in the beam domain during a given OFDM transmission block (where the block index is omitted for brevity) in the DL, then the transmitted signal, $\upnot{\barbx}{dl}\parenth{t}\in\bbC^{M\times1}$, can be represented as \cite{Hwang09OFDM}
\begin{align}\label{eq:bm_ofdm_tx}
\upnot{\barbx}{dl}\parenth{t}
=\sum_{n=0}^{\dnnot{N}{us}-1}\upnot{\barbx}{dl}_{n}\cdot\expx{\barjmath2\pi\frac{n}{\dnnot{T}{us}}t},\ -\dnnot{T}{cp}\leq t<\dnnot{T}{us},
\end{align}
and the DL beam domain signal received by UT $u$ at time $t$ during the given transmission block (in the absence of noise and possible interblock interference for clarity) can be expressed as
\begin{align}\label{eq:rec_bm_dl}
\upnot{\barby_{u}}{dl}\parenth{t}
  &=\int\limits_{-\infty}^{\infty}\!\upnot{\barbG_{u}}{dl}\parenth{t,\tau}\cdot\upnot{\barbx}{dl}\parenth{t-\tau}\intdx{\tau}\in\bbC^{K\times 1},
\end{align}
where $\barbGdl_{u}\parenth{t,\tau}$ is the DL beam domain channel impulse response matrix of UT $u$ given in \eqref{eq:beam_cha_ele_appr_del}.
In this work we focus on the beam domain transmission and adopt the transmission model in \eqref{eq:rec_bm_dl} for clarity. Note that the DL beam domain transmission model in \eqref{eq:rec_bm_dl} can be directly transformed into the space domain using the unitary equivalence between the beam domain and space domain channels.

With the above transmission model, we proceed to investigate the spreading properties of the received signals caused by channel dispersion. From \eqref{eq:beam_cha_ele_appr_del} and \eqref{eq:rec_bm_dl}, the received signal over beam $k$ of UT $u$ at time $t$ is given by
\begin{align}\label{eq:rec_bm_dl_per}
\vecele{\upnot{\barby_{u}}{dl}\parenth{t}}{k}
&=\sum_{m=0}^{M-1}\int\limits_{-\infty}^{\infty}\!
\vecele{\upnot{\barbG_{u}}{dl}\parenth{t,\tau}}{k,m}\cdot\vecele{\upnot{\barbx}{dl}\parenth{t-\tau}}{m}\intdx{\tau}
\ntb
&=\sum_{m=0}^{M-1}\int\limits_{\theta_m}^{\theta_{m+1}}\int\limits_{\phi_k}^{\phi_{k+1}}\!
\sqrt{\ttS_{u}\left(\phi,\theta\right)}
\cdot\expx{\barjmath\zetadl\parenth{\phi,\theta}}
\ntb
&\qquad\cdot\expx{\barjmath2\pi t\nu_u\left(\phi\right)}
\cdot\vecele{\upnot{\barbx}{dl}\parenth{t-\tau_u\left(\phi,\theta\right)}}{m}\intdx{\phi}\intdx{\theta}.
\end{align}
Then the received beam domain signal, $\vecele{\upnot{\barby_{u}}{dl}\parenth{t}}{k}$, will experience time offsets (delays) relative to the transmitted signal, $\upnot{\barbx}{dl}\parenth{t}$, ranging from $\upnot{\tau}{min}_{u,k}$ to $\upnot{\tau}{max}_{u,k}$ \cite{Gallager08Principles}, where
\begin{align}
  \upnot{\tau}{min}_{u,k}&=
\mathop{\min}\limits_{m}
\mathop{\min}\limits_{\substack{\phi\in[\phi_k,\phi_{k+1}]\\ \theta\in[\theta_m,\theta_{m+1}]}}\ \tau_u\left(\phi,\theta\right)
=\mathop{\min}\limits_{\substack{\phi\in[\phi_k,\phi_{k+1}]\\ \theta\in[\theta_0,\theta_{M}]}}\ \tau_u\left(\phi,\theta\right),
\label{eq:min_beam_comp_del}
\\
\upnot{\tau}{max}_{u,k}&=
\mathop{\max}\limits_{m}
\mathop{\max}\limits_{\substack{\phi\in[\phi_k,\phi_{k+1}]\\ \theta\in[\theta_m,\theta_{m+1}]}}\ \tau_u\left(\phi,\theta\right)
=
\mathop{\max}\limits_{\substack{\phi\in[\phi_k,\phi_{k+1}]\\ \theta\in[\theta_0,\theta_{M}]}}\ \tau_u\left(\phi,\theta\right).
\label{eq:max_beam_comp_del}
\end{align}
Meanwhile, $\vecele{\upnot{\barby_{u}}{dl}\parenth{t}}{k}$ will experience frequency offsets relative to the transmitted signal, $\upnot{\barbx}{dl}\parenth{t}$, ranging from $\upnot{\nu}{min}_{u,k}$ to $\upnot{\nu}{max}_{u,k}$ \cite{Gallager08Principles}, where
\begin{align}
  \upnot{\nu}{min}_{u,k}&=
\mathop{\min}\limits_{\phi\in[\phi_k,\phi_{k+1}]}\ \nu_u\left(\phi\right)
=\left(\frac{2k}{K}-1\right)\nu_u,
\label{eq:min_beam_comp_dop}
\\
\upnot{\nu}{max}_{u,k}&=
\mathop{\max}\limits_{\phi\in[\phi_k,\phi_{k+1}]}\ \nu_u\left(\phi\right)
=\left(\frac{2\left(k+1\right)}{K}-1\right)\nu_u,
\label{eq:max_beam_comp_dop}
\end{align}
where the definitions in \eqref{eq:dop_jakes} and \eqref{eq:vir_ang} are used.
For notation simplicity, we denote the minimum and the maximum time and frequency offsets of UT $u$ across all receive beams as
\begin{align}
\upnot{\tau}{min}_{u}&=\min_{k}\left\{\upnot{\tau}{min}_{u,k}\right\},
\label{eq:min_beam_comp_del_all}
\\
\upnot{\tau}{max}_{u}&=\max_{k}\left\{\upnot{\tau}{max}_{u,k}\right\},
\label{eq:max_beam_comp_del_all}
\\
\upnot{\nu}{min}_{u}&=\min_{k}\left\{\upnot{\nu}{min}_{u,k}\right\}=-\nu_u,
\label{eq:min_beam_comp_dop_all}
\\
\upnot{\nu}{max}_{u}&=\max_{k}\left\{\upnot{\nu}{max}_{u,k}\right\}=\nu_u.
\label{eq:max_beam_comp_dop_all}
\end{align}

\subsection{DL Synchronization}

As the performance of OFDM-based transmission is sensitive to time and frequency offsets, it is necessary to perform time and frequency synchronization to compensate for time and frequency offsets of the received signals.
In particular, the received signals should be carefully adjusted so that the resultant minimum time offset and center frequency offset are aligned to zero \cite{Hwang09OFDM}.
The most common synchronization approach for MIMO systems is to compensate for the time and frequency offsets of the received signals in the space domain using the same time and frequency adjustment parameters. Specifically, with time adjustment $\upnot{\tau}{syn}_{u}=\upnot{\tau}{min}_{u}$ and frequency adjustment $\upnot{\nu}{syn}_{u}=\left(\upnot{\nu}{min}_{u}+\upnot{\nu}{max}_{u}\right)/2$ applied to the received space domain signal vector, the resultant beam domain signal is given by
\begin{align}\label{eq:mul_sync}
\upnot{\barby_{u}}{dl,joi}\parenth{t}
&=\upnot{\barby_{u}}{dl}\parenth{t+\upnot{\tau}{syn}_{u}}\cdot\expx{-\barjmath2\pi\left(t+\upnot{\tau}{syn}_{u}\right)\upnot{\nu}{syn}_{u}}.
\end{align}
Then the effective delay and frequency spreads of the adjusted signal, $\upnot{\barby_{u}}{dl,joi}\parenth{t}$, relative to the transmitted signal, $\upnot{\barbx}{dl}\parenth{t}$, are given by
\begin{align}
\upnot{\Delta}{joi}_{\tau_u}
&=\upnot{\tau}{max}_{u}
  -\upnot{\tau}{min}_{u},
  \label{eq:joi_del_spr}\\
\upnot{\Delta}{joi}_{\nu_u}
&=\frac{\upnot{\nu}{max}_{u}-\upnot{\nu}{min}_{u}}{2}
=\nu_u
=\dnnot{f}{c}\frac{v_u}{c},
  \label{eq:joi_dop_spr}
\end{align}
respectively. The terms, $\upnot{\Delta}{joi}_{\tau_u}$ and $\upnot{\Delta}{joi}_{\nu_u}$, are usually referred to as the effective channel delay and frequency spreads \cite{Tse05Fundamentals,Gallager08Principles}, and have great impacts on the design of practical OFDM-based wireless systems. Specifically, the CP length and the OFDM symbol length should be carefully chosen to satisfy $\max_{u}\left\{\upnot{\Delta}{joi}_{\tau_u}\right\}\leq\dnnot{T}{cp}\leq\dnnot{T}{us}\ll 1/\max_{u}\left\{\upnot{\Delta}{joi}_{\nu_u}\right\}$ \cite{Hwang09OFDM}.

From \eqref{eq:joi_dop_spr}, the effective channel frequency spread, $\upnot{\Delta}{joi}_{\nu_u}$, scales linearly with the carrier frequency, $\dnnot{f}{c}$, for a given mobile velocity $v_u$. Therefore, in order to support the same UT mobility, the length of the OFDM symbol in mmW/THz systems would be substantially reduced compared with that in the conventional microwave systems. Meanwhile, the length of the CP would be the same as that in the conventional microwave systems to deal with the same delay spread, which might lead to difficulty in selecting proper OFDM parameters.

Recalling \eqref{eq:min_beam_comp_dop} and \eqref{eq:max_beam_comp_dop}, we can observe that Doppler frequency offsets $\upnot{\nu}{min}_{u,k}$ and $\upnot{\nu}{max}_{u,k}$ for the signals over a particular beam $k$ may be much different from $\upnot{\nu}{min}_{u}$ and $\upnot{\nu}{max}_{u}$ defined in \eqref{eq:min_beam_comp_dop_all} and \eqref{eq:max_beam_comp_dop_all}, so as the time offsets $\upnot{\tau}{min}_{u,k}$ and $\upnot{\tau}{max}_{u,k}$. If these offsets are properly adjusted over each beam individually, the effective delay and frequency spreads of the signals combined from all receive beams can be reduced. Motivated by this, we propose PBS in time and frequency, where adjustment of time and frequency offsets is applied to the signal over each receive beam individually, as detailed below.

Recall the received signal over the \ith{k} beam, namely $\vecele{\upnot{\barby_{u}}{dl}\parenth{t}}{k}$ in \eqref{eq:rec_bm_dl_per}. With time adjustment $\upnot{\tau}{syn}_{u,k}=\upnot{\tau}{min}_{u,k}$ and frequency adjustment $\upnot{\nu}{syn}_{u,k}=\left(\upnot{\nu}{min}_{u,k}+\upnot{\nu}{max}_{u,k}\right)/2$ applied,\footnote{Note that the time and frequency adjustment parameters $\upnot{\tau}{syn}_{u,k}$ and $\upnot{\nu}{syn}_{u,k}$ depend on the long term statistical channel parameters and vary relatively slowly, and thus can be obtained with properly designed synchronization signals \cite{Morelli07Synchronization,Meng16Omnidirectional}.} the adjusted signal is given by
\begin{align}\label{eq:adj_per}
\upnot{\bary_{u,k}}{dl,per}\parenth{t}
&=\vecele{\upnot{\barby_{u}}{dl}\parenth{t+\upnot{\tau}{syn}_{u,k}}}{k}\cdot\expx{-\barjmath2\pi\left(t+\upnot{\tau}{syn}_{u,k}\right)\upnot{\nu}{syn}_{u,k}}.
\end{align}

Combine the adjusted signals over different beams into a vector as
\begin{align}
\upnot{\barby_{u}}{dl,per}\parenth{t}
=\left[\upnot{\bary_{u,0}}{dl,per}\parenth{t}\quad\upnot{\bary_{u,1}}{dl,per}\parenth{t}
\quad\ldots\quad\upnot{\bary_{u,K-1}}{dl,per}\parenth{t}\right]^{T}\in\bbC^{K\times1}.
\end{align}
Then the effective delay and frequency spreads of the adjusted signal, $\upnot{\barby_{u}}{dl,per}\parenth{t}$, relative to the transmitted signal, $\upnot{\barbx}{dl}\parenth{t}$, are given by
\begin{align}
\upnot{\Delta}{per}_{\tau_u}
&=\mathop{\max}\limits_{k}\
\left\{\upnot{\tau}{max}_{u,k}
  -\upnot{\tau}{min}_{u,k}\right\},
  \label{eq:per_del_spr}\\
\upnot{\Delta}{per}_{\nu_u}
&=\mathop{\max}\limits_{k}\
\left\{\frac{\upnot{\nu}{max}_{u,k}-\upnot{\nu}{min}_{u,k}}{2}\right\}
\equaa\frac{\nu_u}{K},
  \label{eq:per_dop_spr}
\end{align}
respectively, where (a) follows from \eqref{eq:min_beam_comp_dop} and \eqref{eq:max_beam_comp_dop}.
The following proposition on the effective channel delay and frequency spreads with PBS can be readily obtained from \eqref{eq:joi_del_spr}, \eqref{eq:joi_dop_spr}, \eqref{eq:per_del_spr}, and \eqref{eq:per_dop_spr}.
\begin{prop}\label{prop:beam_spr_com}
The delay spread, $\upnot{\Delta}{per}_{\tau_u}$, and the frequency spread, $\upnot{\Delta}{per}_{\nu_u}$, of the effective channel with PBS in time and frequency satisfy
\begin{align}\label{eq:beam_comp_spr}
\upnot{\Delta}{per}_{\tau_u}\leq\upnot{\Delta}{joi}_{\tau_u},
\ \
\upnot{\Delta}{per}_{\nu_u}=\frac{\upnot{\Delta}{joi}_{\nu_u}}{K}.
\end{align}
\end{prop}

From \propref{prop:beam_spr_com}, compared with the conventional synchronization approach in \eqref{eq:mul_sync}, the effective channel delay and frequency spreads can be reduced with the proposed PBS approach.
In particular, the effective channel frequency spread is approximately reduced by a factor of the number of UT antennas, $K$, in the large array regime. In addition, the effective channel delay spread can also be reduced with PBS, but the quantitative result is difficult to establish without explicit physical modeling of the propagation delay function $\tau_{u}\parenth{\phi,\theta}$.

However, with the clustering nature of mmW/THz channels taken into account \cite{Akdeniz14Millimeter,Adhikary14Joint-jsac}, a significant reduction in the effective channel delay spread can be still expected.
To provide some insights on the reduction in delay spread with PBS, we herein consider a special but important case in which a ring of scatterers are located around UTs \cite{Shiu00Fading,Patzold06wideband}. Assume that the radius of the ring of the scatterers around UT $u$ is $r_u$, then the propagation delay of the channel path with the AoA, $\phi$, is given by $\upnot{\tau_{u}}{oner}\parenth{\phi,\theta}\triangleq r_u/c\times\left[1+\sin\parenth{\phi}\right]$ \cite{Patzold06wideband}. From \eqref{eq:joi_del_spr} and \eqref{eq:per_del_spr}, the effective channel delay spreads with the conventional synchronization and PBS are given by
\begin{align}
  \upnot{\Delta_{\tau_u}}{oner,joi}=\frac{2r_u}{c},\ \
  \upnot{\Delta_{\tau_u}}{oner,per}=\frac{2r_u}{Kc}=\frac{\upnot{\Delta_{\tau_u}}{oner,joi}}{K},\label{eq:del_oner_comp}
\end{align}
respectively. Thus, for the one-ring case, the effective channel delay spread with PBS is also reduced by a factor of the number of UT antennas, $K$.

The result in \propref{prop:beam_spr_com} can be exploited to simplify the implementation and improve the performance of mmW/THz massive MIMO-OFDM systems. In particular, the number of UT antennas, $K$, also scales linearly with the carrier frequency for the same antenna array aperture although the maximum channel Doppler shift, $\nu_u$, scales linearly with the carrier frequency. Thus, assuming a fixed antenna array aperture, the effective channel Doppler frequency spread over mmW/THz bands becomes approximately the same as that over regular bands with PBS, which can mitigate severe Doppler effects over mmW/THz channels. Moreover, the effective channel delay spread can be significantly reduced with PBS, which can further lead to a substantial reduction in the CP overhead.

With PBS presented above, the CP length and the OFDM symbol length can be chosen to satisfy
$\max_{u}\left\{\upnot{\Delta}{per}_{\tau_u}\right\}\leq\dnnot{T}{cp}\leq\dnnot{T}{us}\ll 1/\max_{u}\left\{\upnot{\Delta}{per}_{\nu_u}\right\}$ for the mmW/THz systems even in high mobility scenarios.
Then the demodulated OFDM symbol over beam $k$ of UT $u$ at subcarrier $n$ in the given block is given by \cite{Hwang09OFDM}
\begin{align}\label{eq:beam_fre_cha_eff_trans}
\vecele{\upnot{\barby_{u,n}}{dl}}{k}
&=\oneon{\dnnot{T}{us}}\int\limits_{0}^{\dnnot{T}{us}}\!
\vecele{\upnot{\barby_{u}}{dl,per}\parenth{t}}{k}
\cdot\expx{-\barjmath2\pi\frac{n}{\dnnot{T}{us}}t}\intdx{t}
\ntb
&\equaa\oneon{\dnnot{T}{us}}\int\limits_{0}^{\dnnot{T}{us}}\!
\vecele{\upnot{\barby_{u}}{dl}\parenth{t+\upnot{\tau}{syn}_{u,k}}}{k}
\cdot\expx{-\barjmath2\pi\left(t+\upnot{\tau}{syn}_{u,k}\right)\upnot{\nu}{syn}_{u,k}}
\cdot\expx{-\barjmath2\pi\frac{n}{\dnnot{T}{us}}t}\intdx{t}
\ntb
&\equab\sum_{m=0}^{M-1}\oneon{\dnnot{T}{us}}\int\limits_{0}^{\dnnot{T}{us}}\int\limits_{\theta_m}^{\theta_{m+1}}\int\limits_{\phi_k}^{\phi_{k+1}}\!
\sqrt{\ttS_{u}\left(\phi,\theta\right)}
\cdot\expx{\barjmath\zetadl\parenth{\phi,\theta}}
\ntb
&\qquad
\cdot\expx{\barjmath2\pi\left(t+\upnot{\tau}{syn}_{u,k}\right)\left(\nu_u\parenth{\phi}-\upnot{\nu}{syn}_{u,k}\right)}
\ntb
&\qquad
\cdot\vecele{\upnot{\barbx}{dl}\parenth{t-\left(\tau_u\left(\phi,\theta\right)-\upnot{\tau}{syn}_{u,k}\right)}}{m}
\cdot\expx{-\barjmath2\pi\frac{n}{\dnnot{T}{us}}t}
\intdx{\phi}\intdx{\theta}\intdx{t}
\ntb
&\mathop{\simeq}^{\mathrm{\left(c\right)}}
\sum_{m=0}^{M-1}\oneon{\dnnot{T}{us}}\int\limits_{0}^{\dnnot{T}{us}}\int\limits_{\theta_m}^{\theta_{m+1}}\int\limits_{\phi_k}^{\phi_{k+1}}\!
\sqrt{\ttS_{u}\left(\phi,\theta\right)}
\cdot\expx{\barjmath\zetadl\parenth{\phi,\theta}}
\ntb
&\qquad
\cdot\expx{\barjmath2\pi\upnot{\tau}{syn}_{u,k}\left(\nu_u\parenth{\phi}-\upnot{\nu}{syn}_{u,k}\right)}
\ntb
&\qquad
\cdot\vecele{\upnot{\barbx}{dl}\parenth{t-\left(\tau_u\left(\phi,\theta\right)-\upnot{\tau}{syn}_{u,k}\right)}}{m}
\cdot\expx{-\barjmath2\pi\frac{n}{\dnnot{T}{us}}t}
\intdx{\phi}\intdx{\theta}\intdx{t}
\ntb
&\equad\sum_{m=0}^{M-1}\vecele{\upnot{\barbG}{dl,per}_{u,n}}{k,m}\vecele{\upnot{\barbx}{dl}_{n}}{m},
\end{align}
where (a) follows from \eqref{eq:adj_per}, (b) follows from \eqref{eq:rec_bm_dl_per},
the approximation in (c) follows from $t\left(\nu_u\parenth{\phi}-\upnot{\nu}{syn}_{u,k}\right)\ll1$ for $0\leq t\leq\dnnot{T}{us}$,
(d) follows from \eqref{eq:bm_ofdm_tx}, and
$\upnot{\barbG}{dl,per}_{u,n}$ denotes the frequency response of the effective DL beam domain channel with PBS between the BS and UT $u$ at subcarrier $n$ during the given transmission block given by
\begin{align}\label{eq:beam_fre_cha_eff}
\vecele{\upnot{\barbG}{dl,per}_{u,n}}{k,m}
&\triangleq
\int\limits_{\theta_m}^{\theta_{m+1}}\int\limits_{\phi_k}^{\phi_{k+1}}\!
\sqrt{\ttS_{u}\left(\phi,\theta\right)}
\cdot\expx{\barjmath\zetadl\parenth{\phi,\theta}}
\cdot\expx{\barjmath2\pi\upnot{\tau}{syn}_{u,k}\left(\nu_u\parenth{\phi}-\upnot{\nu}{syn}_{u,k}\right)}
\ntb
&\qquad\cdot\expx{-\barjmath2\pi\frac{n}{\dnnot{T}{us}}\left(\tau_u\left(\phi,\theta\right)-\upnot{\tau}{syn}_{u,k}\right)}
\intdx{\phi}\intdx{\theta}.
\end{align}

Thus, the DL beam domain transmission model for mmW/THz massive MIMO-OFDM can be represented in a concise per-subcarrier manner as
\begin{align}\label{eq:beam_fre_tra_model}
\upnot{\barby_{u,n}}{dl}=\upnot{\barbG}{dl,per}_{u,n}\barbxdl_{n}\in\bbC^{K\times 1},\qquad n=0,1,\ldots,\dnnot{N}{us}-1.
\end{align}
Note that if the conventional synchronization approach in \eqref{eq:mul_sync} is adopted, it would be difficult to choose the CP length and the OFDM symbol length to satisfy the previously mentioned wireless OFDM design requirements in the considered mmW/THz systems. In such scenarios, a complicated transmission model involving intercarrier interference and/or interblock interference should be considered \cite{Hwang09OFDM}.

\subsection{UL Synchronization}

In the above we focus on PBS for the DL. Now we address the UL case via leveraging the reciprocity of the UL and DL physical parameters.
Let $\left\{\upnot{\barbx}{ul}_{u,n}\right\}_{n=0}^{\dnnot{N}{us}-1}$ be the complex-valued symbols to be transmitted by UT $u$ in the beam domain during a given OFDM block, then the transmitted signal, $\upnot{\barbx}{ul}_{u}\parenth{t}\in\bbC^{K\times1}$, can be represented as
\begin{align}\label{eq:bm_ofdm_tx_ul}
\upnot{\barbx}{ul}_{u}\parenth{t}
=\sum_{n=0}^{\dnnot{N}{us}-1}\upnot{\barbx}{ul}_{u,n}\cdot\expx{\barjmath2\pi\frac{n}{\dnnot{T}{us}}t},\ -\dnnot{T}{cp}\leq t<\dnnot{T}{us}.
\end{align}

As the UL waveform received at the BS is a combination of signals transmitted from different UTs, we propose to perform PBS at the UT sides. In particular, with time adjustment $\upnot{\tau}{syn}_{u,k}=\upnot{\tau}{min}_{u,k}$ and frequency adjustment $\upnot{\nu}{syn}_{u,k}=\left(\upnot{\nu}{min}_{u,k}+\upnot{\nu}{max}_{u,k}\right)/2$ applied to $\vecele{\upnot{\barbx_{u}}{ul}\parenth{t}}{k}$, the adjusted signal is given by
\begin{align}\label{eq:adj_per_ul}
\upnot{\barx_{u,k}}{ul,per}\parenth{t}
&=\vecele{\upnot{\barbx_{u}}{ul}\parenth{t+\upnot{\tau}{syn}_{u,k}}}{k}\cdot\expx{-\barjmath2\pi\left(t+\upnot{\tau}{syn}_{u,k}\right)\upnot{\nu}{syn}_{u,k}}.
\end{align}
Then the beam domain signal received at the BS at time $t$ during the given transmission block (in the absence of noise for clarity) can be represented as
\begin{align}\label{eq:rec_bm_ul}
\vecele{\upnot{\barby}{ul}\parenth{t}}{m}
  &=\sum_{u=0}^{U-1}\sum_{k=0}^{K-1}\int\limits_{-\infty}^{\infty}\!
  \vecele{\upnot{\barbG_{u}}{ul}\parenth{t,\tau}}{m,k}\cdot
  \upnot{\barx_{u,k}}{ul,per}\parenth{t-\tau}\intdx{\tau},
\end{align}
where $\barbGul_{u}\parenth{t,\tau}$ is given in \eqref{eq:beam_cha_ele_appr_del_ul}.

Similarly as the DL case, PBS in the UL can effectively reduce the channel delay and Doppler spreads. Thus, the demodulated OFDM symbols over beam $m$ of the BS at subcarrier $n$ in the given transmission block can be written as \cite{Hwang09OFDM}
\begin{align}
\vecele{\upnot{\barby_{n}}{ul}}{m}
=\sum_{u=0}^{U-1}\sum_{k=0}^{K-1}\vecele{\upnot{\barbG}{ul,per}_{u,n}}{m,k}\vecele{\barbxul_{u,n}}{k},
\end{align}
where $\upnot{\barbG}{ul,per}_{u,n}$ denotes the frequency response of the effective UL beam domain channel between the BS and UT $u$ at subcarrier $n$ given by
\begin{align}\label{eq:beam_fre_cha_eff_ul}
\vecele{\upnot{\barbG}{ul,per}_{u,n}}{m,k}
&\triangleq
\int\limits_{\theta_m}^{\theta_{m+1}}\int\limits_{\phi_k}^{\phi_{k+1}}\!
\sqrt{\ttS_{u}\left(\phi,\theta\right)}
\cdot\expx{\barjmath\zetaul\parenth{\phi,\theta}}
\cdot\expx{\barjmath2\pi\upnot{\tau}{syn}_{u,k}\left(\nu_u\parenth{\phi}-\upnot{\nu}{syn}_{u,k}\right)}
\ntb
&\qquad\cdot\expx{-\barjmath2\pi\frac{n}{\dnnot{T}{us}}\left(\tau_u\left(\phi,\theta\right)-\upnot{\tau}{syn}_{u,k}\right)}
\intdx{\phi}\intdx{\theta}.
\end{align}
Then the UL beam domain transmission model for mmW/THz massive MIMO-OFDM can be represented in a per-subcarrier manner as
\begin{align}\label{eq:beam_fre_tra_model_ul}
\upnot{\barby_{n}}{ul}=\sum_{u=0}^{U-1}\upnot{\barbG}{ul,per}_{u,n}\barbxul_{u,n}\in\bbC^{M\times 1},\qquad n=0,1,\ldots,\dnnot{N}{us}-1.
\end{align}

\subsection{Discrete Time Channel Statistics}

Statistical properties of the discrete time beam domain channels can be similarly derived. From \eqref{eq:beam_fre_cha_eff} and \eqref{eq:beam_fre_cha_eff_ul}, the beam domain channel elements are uncorrelated in the sense that
\begin{align}
\expect{\vecele{\upnot{\barbG}{dl,per}_{u,n}}{k,m}\vecele{\upnot{\barbG}{dl,per}_{u,n}}{k',m'}^{*}}
&=\vecele{\bOmega_u}{k,m}\cdot\delfunc{k-k'}\delfunc{m-m'},\label{eq:beam_cha_dis_unc}\\
\expect{\vecele{\upnot{\barbG}{ul,per}_{u,n}}{m,k}\vecele{\upnot{\barbG}{ul,per}_{u,n}}{m',k'}^{*}}
&=\vecele{\bOmega_u}{k,m}\cdot\delfunc{k-k'}\delfunc{m-m'},\label{eq:beam_cha_dis_unc_ul}
\end{align}
where $\bOmega_u$ is the beam domain channel power matrix defined in \eqref{eq:beam_cha_pow}.
According to the law of large numbers, the beam domain channel elements exhibit a Gaussian distribution, i.e., $\vecele{\upnot{\barbG}{dl,per}_{u,n}}{k,m}\sim\GCN{0}{\vecele{\bOmega_u}{k,m}}$ and $\vecele{\upnot{\barbG}{ul,per}_{u,n}}{m,k}\sim\GCN{0}{\vecele{\bOmega_u}{k,m}}$.
We define the average squared channel norms of beam $m$ at the BS side and beam $k$ at UT $u$ side as
\begin{align}
\upnot{\omega}{bs}_{u,m}
&\triangleq\sum_{k=0}^{K-1}\vecele{\bOmega_{u}}{k,m},\ m=0,1,\dots,M-1 \label{eq:tra_bm_gain},\\
\upnot{\omega}{ut}_{u,k}
&\triangleq\sum_{m=0}^{M-1}\vecele{\bOmega_{u}}{k,m},\ k=0,1,\dots,K-1,\label{eq:tra_bm_gain_ul_tra}
\end{align}
respectively, which will be useful for transmission design investigated in the following section.

\section{BDMA with PBS}\label{sec:bdma_pbs_syn}

With PBS proposed above, the effective channel frequency spread in the beam domain over mmW/THz bands becomes almost the same as that over regular wireless bands meanwhile the effective channel delay spread in the beam domain can be significantly reduced. The proposed PBS can be embedded into all mmW/THz massive MIMO transmissions.

BDMA in \cite{Sun15Beam} is an attractive approach for mmW/THz massive MIMO particularly in high mobility scenarios for the following reasons. First, beam domain channels at mmW/THz bands exhibit an approximately sparse nature \cite{Heath16overview,Gao17Fast} and therefore, BDMA is well suited for such channels \cite{Sun15Beam}.
Second, transmitters only need to know the statistical channel state information (CSI), which avoids the challenge in acquisition of the instantaneous CSI required for conventional massive MIMO transmission over mmW/THz channels \cite{Heath16overview} and is attractive for transmission in high mobility scenarios \cite{Akdeniz14Millimeter}. Third, the implementation complexity of BDMA is relatively low as only beam scheduling and power allocation for different UTs based on the beam domain channel statistics are required instead of complicated multiuser precoding and detection. In this section, we will investigate BDMA with PBS for mmW/THz massive MIMO.

\subsection{DL BDMA}

We first outline BDMA for DL massive MIMO transmission in \cite{Sun15Beam}.
From \eqref{eq:beam_fre_tra_model}, the DL beam domain transmission model can be rewritten as
\begin{align}\label{eq:bdma_dl_bm_re}
\upnot{\barby_{u}}{dl}=\upnot{\barbG}{dl,per}_{u}\barbxdl_{u}+\upnot{\barbG}{dl,per}_{u}\sum_{u'\neq u}\barbxdl_{u'}+\barbzdl_{u}\in\bbC^{K\times 1},
\end{align}
where the subcarrier index is omitted for brevity, $\barbzdl_{u}$ is the effective DL noise distributed as $\GCN{\bzero}{\upnot{\sigma}{dl}\bI_{K}}$, and $\barbxdl_{u}$ is the DL beam domain transmitted signal for UT $u$. We assume that the signals intended for different UTs are uncorrelated, and denote $\upnot{\barbQ_{u}}{dl}=\expect{\upnot{\barbx_{u}}{dl}\left(\upnot{\barbx_{u}}{dl}\right)^H}\in\bbC^{M\times M}$ as the beam domain transmit covariance of UT $u$.

With the assumption that each UT knows its instantaneous DL CSI\footnote{The effective channel Doppler spread is significantly reduced with PBS, and the instantaneous DL CSI can be obtained by the UTs through properly designed DL pilot signals \cite{Sun15Beam}.} and the BS only knows the statistical CSI of all UTs, the DL ergodic achievable sum rate is given by
\begin{align}\label{eq:erg_rate_dl}
\upnot{R}{dl}
  &=\sum_{u=0}^{U-1}\Exp\Bigg\{\log_2\det\left(\upnot{\sigma}{dl}\bI+\sum_{u'=0}^{U-1}\upnot{\barbG_{u}}{dl,per}\updl{\barbQ_{u'}}\left(\upnot{\barbG_{u}}{dl,per}\right)^H\right)\ntb
&\qquad\qquad-\log_2\det\left(\upnot{\sigma}{dl}\bI+\sum_{u'\neq u}\upnot{\barbG_{u}}{dl,per}\updl{\barbQ_{u'}}\left(\upnot{\barbG_{u}}{dl,per}\right)^H\right)
\Bigg\},
\end{align}
where the expectation is with respect to the channel realizations \cite{Sun15Beam}.
With the sum rate expression in \eqref{eq:erg_rate_dl} and the uncorrelated properties of the beam domain channel elements in \eqref{eq:beam_cha_dis_unc}, the structures of the DL transmit covariances that can maximize $\upnot{R}{dl}$ have been investigated in \cite{Sun15Beam}. Specifically, denote the eigenvalue decomposition of the transmit covariance as $\barbQdl_u=\upnot{\barbU}{dl}_u\diag{\upnot{\blambda}{dl}_u}\left(\upnot{\barbU}{dl}_u\right)^{H}$, where the columns of $\upnot{\barbU}{dl}_u$ are the eigenvectors of $\barbQdl_u$ and the entries of $\upnot{\blambda}{dl}_u$ are the eigenvalues of $\barbQdl_u$, then the DL beam domain transmit covariances satisfy the following structures:
\begin{align}\label{eq:opt_cond_bm}
  \upnot{\barbU}{dl}_u&=\bI,\ \forall u ,\\
\left(\upnot{\blambda}{dl}_u\right)^{T}\upnot{\blambda}{dl}_{u'}&=0,\ \forall u\neq u'.
\end{align}

The above structures of the DL transmit covariance matrices have immediate engineering meaning. Specifically, $\upnot{\barbU}{dl}_u=\bI$ indicates that the DL signals should be transmitted in the beam domain. Meanwhile, $\left(\upnot{\blambda}{dl}_u\right)^{T}\upnot{\blambda}{dl}_{u'}=0$ for $u\neq u'$ indicates that one DL transmit beam can be allocated to at most one UT.
Thus, finding the DL beam domain transmit covariance matrices is equivalent to scheduling non-overlapping transmit beam sets for different UTs and properly performing power allocation across different scheduled transmit beams. As equal power allocation across scheduled subchannels usually has a near-optimal performance \cite{Chen08Multimode}, we therefore focus on beam scheduling for different UTs.

Based on the above DL transmit covariance structures, BDMA, in which multiple access is realized by providing each UT with a mutually non-overlapping BS beam set, has been proposed in \cite{Sun15Beam}. Now we investigate DL beam scheduling for different UTs.
Denote $\upnot{\cB_{u}}{dl,bs}$ and $\upnot{\cB_{u}}{dl,ut}$ as the DL transmit and receive beam sets scheduled for UT $u$, respectively, then the DL ergodic achievable sum rate in \eqref{eq:erg_rate_dl} with equal power allocation is given by
\begin{align}\label{eq:rate_cpa_dl}
\upnot{R}{dl,epa}
=\sum_{u=0}^{U-1}\expect{\log_2
\frac{\det\left(\bI+\frac{\upnot{\rho}{dl}}{\sum_{u'=0}^{U-1}\abs{\upnot{\cB_{u'}}{dl,bs}}}
\sum_{u''=0}^{U-1}
\vecele{\upnot{\barbG_{u}}{dl,per}}{\upnot{\cB_u}{dl,ut},\upnot{\cB_{u''}}{dl,bs}}\vecele{\upnot{\barbG_{u}}{dl,per}}{\upnot{\cB_u}{dl,ut},\upnot{\cB_{u''}}{dl,bs}}^H\right)}
{\det\left(\bI+\frac{\upnot{\rho}{dl}}{\sum_{u'=0}^{U-1}\abs{\upnot{\cB_{u'}}{dl,bs}}}
\sum_{u''\neq u}
\vecele{\upnot{\barbG_{u}}{dl,per}}{\upnot{\cB_u}{dl,ut},\upnot{\cB_{u''}}{dl,bs}}\vecele{\upnot{\barbG_{u}}{dl,per}}{\upnot{\cB_u}{dl,ut},\upnot{\cB_{u''}}{dl,bs}}^H\right)}
},
\end{align}
where $\upnot{\rho}{dl}=\upnot{P}{dl}/\upnot{\sigma}{dl}$ is the DL signal-to-noise ratio (SNR) and $\upnot{P}{dl}$ is the DL sum power budget.

The DL beam scheduling problem can be formulated as follows:
\begin{subequations}\label{eq:pa_dl_besc}
\begin{align}
  \maxi{\left\{\upnot{\cB}{dl,bs}_u,\upnot{\cB}{dl,ut}_u:u\in\cU\right\}}\quad &\upnot{R}{dl,epa},
\\
\st\quad
& \upnot{\cB_{u}}{dl,bs}\cap\upnot{\cB_{u'}}{dl,bs}=\varnothing,\ \forall u\neq u', \label{eq:bl_consa}\\
& \abs{\upnot{\cB_{u}}{dl,bs}}\leq \upnot{B_u}{dl,bs},\ \forall u, \label{eq:bl_consb}\\
& \abs{\upnot{\cB_{u}}{dl,ut}}\leq \upnot{B_u}{dl,ut},\ \forall u, \label{eq:bl_consc}\\
& \sum_{u=0}^{U-1}\abs{\upnot{\cB_{u}}{dl,bs}}\leq \upnot{B}{dl,bs},\label{eq:bl_consd}
\end{align}
\end{subequations}
where $\upnot{B_u}{dl,bs}$, $\upnot{B_u}{dl,ut}$, and $\upnot{B}{dl,bs}$ are the maximum allowable numbers of transmit, receive beams for UT $u$, and total transmit beams in the DL, respectively. Note that the numbers of maximum allowable beams can be adjusted to control the required numbers of radio frequency chains in mmW/THz massive MIMO.

The optimization problem in \eqref{eq:pa_dl_besc} is in general difficult due to the stochastic nature of the objective function $\upnot{R}{dl,epa}$ in \eqref{eq:erg_rate_dl} and the combinatorial nature of beam scheduling, especially for the considered mmW/THz massive MIMO systems with large numbers of antennas and UTs, and the optimal solution must be found through an exhaustive search.
In order to obtain a feasible solution of \eqref{eq:pa_dl_besc} with relatively low complexity, we provide here
a (suboptimal) norm-based DL greedy beam scheduling algorithm motivated by \cite{Sun15Beam}.
In particular, the BS first schedules the DL transmit beams for different UTs with all receive beams temporarily activated based on the ordering of the average squared beam domain channel norm at the BS side, $\upnot{\omega}{bs}_{u,m}$, defined in \eqref{eq:tra_bm_gain}, and then schedules receive beams of different UTs based on the ordering of the average squared beam domain channel norm at the UT side, $\upnot{\omega}{ut}_{u,k}$, defined in \eqref{eq:tra_bm_gain_ul_tra}. The description of the DL greedy beam scheduling algorithm is summarized in \tabref{tab:GBSA}.

\begin{table}[!t]
\caption{DL Greedy Beam Scheduling Algorithm}\label{tab:GBSA}
\centering
\vspace{-0.8cm}
\begin{algorithm}[H]
\label{alg:GBSA}
\begin{algorithmic}[1]
\Require
The UT set $\cU$ and the beam domain channel power matrices $\left\{\bOmega_{u}:u\in\cU\right\}$
\Ensure
DL beam scheduling pattern $\left\{\upnot{\cB_{u}}{dl,bs},\upnot{\cB_{u}}{dl,ut}:u\in\cU\right\}$
\State Initialize $\upnot{\cB_{u}}{dl,bs}=\varnothing$ for all $u$, $\upnot{\cS}{temp}=\varnothing$, and $R=0$
\State Temporarily activate all DL receive beams: Set $\upnot{\cB_{u}}{dl,ut}=\left\{0,1,\ldots,K-1\right\}$ for all $u$
\While{$\abs{\upnot{\cS}{temp}}< MU$}
\State Search for $\left\{\left(u',m'\right)\right\}=\argmax{\left\{\left(u,m\right)\right\}\notin\upnot{\cS}{temp}}\ \upnot{\omega}{bs}_{u,m}$, update $\upnot{\cB_{u'}}{dl,bs}\gets\upnot{\cB_{u'}}{dl,bs}\cup\left\{m'\right\}$, and calculate $\dnnot{R}{temp}=\upnot{R}{dl,epa}$ using \eqref{eq:rate_cpa_dl}
\If{$\dnnot{R}{temp}>R$}
\State Update $R=\dnnot{R}{temp}$
\If{$\sum_{u\in\cU}\abs{\upnot{\cB_{u}}{dl,bs}}\geq \upnot{B}{dl,bs}$}
\State Break
\EndIf
\If{$\abs{\upnot{\cB_{u'}}{dl,bs}}\geq \upnot{B_{u'}}{dl,bs}$}
\State Update $\upnot{\cS}{temp}\gets\upnot{\cS}{temp}\cup\left\{\left(u',m\right)\right\}$ for all $m$
\EndIf
\State Update $\upnot{\cS}{temp}\gets\upnot{\cS}{temp}\cup\left\{\left(u,m'\right)\right\}$ for all $u$
\Else
\State Update $\upnot{\cB_{u'}}{dl,bs}\gets\setdif{\upnot{\cB_{u'}}{dl,bs}}{\left\{m'\right\}}$, and $\upnot{\cS}{temp}\gets\upnot{\cS}{temp}\cup\left\{\left(u',m'\right)\right\}$
\EndIf
\EndWhile
\State Set $\upnot{\cB_{u}}{dl,ut}=\varnothing$ and $\upnot{\cB_u}{uns,ut}=\left\{0,1,\ldots,K-1\right\}$ for all $u$, initialize $u=0$ and $R=0$
\While{$u\leq U-1$}
\State Select receive beam $k'=\argmax{k\in\upnot{\cB_u}{uns,ut}}\ \upnot{\omega}{ut}_{u,k}$, set $\upnot{\cB_u}{uns,ut}\gets\setdif{\upnot{\cB_u}{uns,ut}}{\left\{k'\right\}}$, temporarily update $\upnot{\cB_{u}}{dl,ut}\gets\upnot{\cB_{u}}{dl,ut}\cup\left\{k'\right\}$, and calculate $\dnnot{R}{temp}=\upnot{R}{dl,epa}$ using \eqref{eq:rate_cpa_dl}
\If{$\dnnot{R}{temp}>R$}
\State Update $R=\dnnot{R}{temp}$
\Else
\State Update $\upnot{\cB_{u}}{dl,ut}\gets\setdif{\upnot{\cB_{u}}{dl,ut}}{\left\{k'\right\}}$
\EndIf
\If{$\abs{\upnot{\cB_{u}}{dl,ut}}\geq\upnot{B_u}{dl,ut}$ or $\abs{\upnot{\cB_u}{uns,ut}}\leq 0$}
\State Update $u\gets u+1$
\EndIf
\EndWhile
\end{algorithmic}
\end{algorithm}
\end{table}

\subsection{UL BDMA}

Motivated by the above presented DL BDMA, we consider in this subsection BDMA for UL transmission. In particular, each UT is allocated with a mutually non-overlapping subset of the total receive beams of the BS during the UL. Then UL signal detection for each UT is performed based on the signals received on the allocated receive beams and complicated multiuser detection is not required in UL BDMA.

Denote the BS beam set allocated to UT $u$ in the UL as $\upnot{\cB_u}{ul,bs}$, then from \eqref{eq:beam_fre_tra_model_ul}, the received signal of UT $u$ over the allocated BS beam subsets can be represented as
\begin{align}\label{eq:bdma_ul_bm_re_sud}
\upnot{\barby_{u}}{ul}&=\vecele{\upnot{\barby}{ul}}{\upnot{\cB_u}{ul,bs}}\ntb
&=\vecele{\upnot{\barbG}{ul,per}_{u}}{\upnot{\cB_u}{ul,bs},:}\barbxul_{u}
+\sum_{u'\neq u}
\vecele{\upnot{\barbG}{ul,per}_{u'}}{\upnot{\cB_u}{ul,bs},:}\barbxul_{u'}
+\vecele{\barbzul}{\upnot{\cB_u}{ul,bs}}\in\bbC^{\abs{\upnot{\cB_u}{ul,bs}}\times 1},
\end{align}
where the subcarrier index is omitted for brevity, $\barbzul$ is the UL noise distributed as $\GCN{\bzero}{\upnot{\sigma}{ul}\bI_{M}}$,
and $\barbxul_{u}$ is the UL beam domain transmitted signal of UT $u$.

Similarly to DL BDMA, the transmit directions of all UTs' signals are aligned to the beam domain in UL BDMA, i.e.,
$\expect{\upnot{\barbx_{u}}{ul}\left(\upnot{\barbx_{u}}{ul}\right)^H}$ is diagonal. We assume equal power allocation \cite{Chen08Multimode} across the scheduled transmit beams in the UL and focus on beam scheduling.

With the assumption that the UTs know the statistical CSI of themselves while the BS can access to the instantaneous UL CSI of UTs over the scheduled beams,\footnote{Similar to the DL case, with PBS, the instantaneous UL CSI can be obtained at the BS through properly designed UL pilot signals \cite{Sun15Beam}. Note that the corresponding pilot overhead scales linearly with the number of scheduled transmit beams that is usually much smaller than that of transmit antennas in UL mmW/THz massive MIMO.} the corresponding UL ergodic achievable sum rate with equal power allocation is given by
\begin{align}\label{eq:rate_cpa_ul}
\upnot{R}{ul,epa}=\sum_{u=0}^{U-1}\expect{\log_2
\frac{\det\left(\bI+\sum_{u'=0}^{U-1}\frac{\upul{\rho_{u'}}}{\abs{\upnot{\cB_{u'}}{ul,ut}}}\vecele{\upnot{\barbG}{ul,per}_{u'}}{\upnot{\cB_u}{ul,bs},\upnot{\cB_{u'}}{ul,ut}}
\vecele{\upnot{\barbG}{ul,per}_{u'}}{\upnot{\cB_u}{ul,bs},\upnot{\cB_{u'}}{ul,ut}}^H\right)}
  {\det\left(\bI+\sum_{u'\neq u}\frac{\upul{\rho_{u'}}}{\abs{\upnot{\cB_{u'}}{ul,ut}}}\vecele{\upnot{\barbG}{ul,per}_{u'}}{\upnot{\cB_u}{ul,bs},\upnot{\cB_{u'}}{ul,ut}}
\vecele{\upnot{\barbG}{ul,per}_{u'}}{\upnot{\cB_u}{ul,bs},\upnot{\cB_{u'}}{ul,ut}}^H\right)}},
\end{align}
where $\upnot{\cB_{u}}{ul,ut}$ is the scheduled UL transmit beam set of UT $u$, $\upul{\rho_u}=\upul{P_u}/\upul{\sigma}$ and $\upul{P_u}$ are the UL SNR and power budget of UT $u$, respectively.
Then the UL beam scheduling problem can be formulated as follows:
\begin{subequations}\label{eq:pa_ul_besc}
\begin{align}
  \maxi{\left\{\upnot{\cB}{ul,bs}_u,\upnot{\cB}{ul,ut}_u:u\in\cU\right\}}\quad &\upnot{R}{ul,epa},\\
\st\quad
& \upnot{\cB_{u}}{ul,bs}\cap\upnot{\cB_{u'}}{ul,bs}=\varnothing,\ \forall u\neq u', \label{eq:ul_consa}\\
& \abs{\upnot{\cB_{u}}{ul,bs}}\leq \upnot{B_u}{ul,bs},\ \forall u, \label{eq:ul_consb}\\
& \abs{\upnot{\cB_{u}}{ul,ut}}\leq \upnot{B_u}{ul,ut},\ \forall u, \label{eq:ul_consc}\\
&\sum_{u=0}^{U-1}\abs{\upnot{\cB_{u}}{ul,bs}}\leq \upnot{B}{ul,bs},\label{eq:ul_consd}
\end{align}
\end{subequations}
where $\upnot{B_u}{ul,bs}$, $\upnot{B_u}{ul,ut}$, and $\upnot{B}{ul,bs}$ are the maximum allowable numbers of receive, transmit beams of UT $u$, and total receive beams in the UL, respectively.

The UL beam scheduling problem in \eqref{eq:pa_ul_besc} exhibits a similar structure as the DL problem in \eqref{eq:pa_dl_besc}. Therefore, a (suboptimal) norm-based UL greedy beam scheduling algorithm as the DL case with the objective function correspondingly changed can be similarly developed. The detailed algorithm description is omitted here for brevity.

\section{Simulation Results}\label{sec:pbs_sim}

In this section, simulation results are provided to illustrate the performance of BDMA with PBS at mmW/THz bands. In the simulation, we focus on the DL transmission, and the UL transmission, which exhibits similar results, is omitted here for brevity. Two typical mmW/THz carrier frequencies, 30 GHz and 300 GHz, are considered. The array topology is set as ULA with half wavelength antenna spacing for both the BS and the UT sides. The major MIMO-OFDM parameters are listed in \tabref{tb:ofdm_para_mmw}. Both the maximum allowable numbers of DL transmit and receive beams for each UT are set as 16, and the maximum allowable number of total transmit beams is set to be the same as the number of BS antennas.

\newcolumntype{L}{>{\hspace*{-\tabcolsep}}l}
\newcolumntype{R}{c<{\hspace*{-\tabcolsep}}}
\definecolor{lightblue}{rgb}{0.93,0.95,1.0}
\begin{table}[!t]
\caption{MIMO-OFDM System Parameters}\label{tb:ofdm_para_mmw}
\centering
\ra{1.3}
\begin{tabular}{LccR}
\toprule
Parameter & & \multicolumn{2}{c}{Value}\\
\midrule
\rowcolor{lightblue}
Carrier frequency && 30 GHz & 300 GHz\\
Number of BS antennas $M$ && 128 & 256 \\
\rowcolor{lightblue}
Number of UT antennas $K$ && 32 & 128 \\
System bandwidth && \multicolumn{2}{c}{100 MHz}\\
\rowcolor{lightblue}
Sampling interval $\dnnot{T}{s}$ && \multicolumn{2}{>{\columncolor{lightblue}}c}{6.51 ns}\\
Subcarrier spacing && \multicolumn{2}{c}{75 kHz} \\
\rowcolor{lightblue}
Number of subcarriers $\dnnot{N}{us}$ && \multicolumn{2}{>{\columncolor{lightblue}}c}{2048}\\
CP length $\dnnot{N}{cp}$ && \multicolumn{2}{c}{144}\\
\bottomrule
\end{tabular}
\end{table}

Assume that there are $U=20$ uniformly distributed UTs in a $120^{\circ}$ sector, and the mean channel AoD is uniformly distributed in $[-\pi/3,\pi/3]$ in radians. All UTs are assumed to be at the same distance from the BS, and the path loss is set as unit.
The random channel realizations are generated using a similar procedure as the WINNER II channel model \cite{win2chanmod}, which has been widely adopted in mmW/THz related works \cite{Lin16Energy,Akdeniz14Millimeter}. The number of channel clusters is set as 4 and each of the clusters is composed of 20 subpaths \cite{Akdeniz14Millimeter}. The delay spread and angle spread are set as 1388.4 ns and $2^{\circ}$, respectively \cite{Rappaport13Millimeter}.

\begin{figure*}[!t]
\centering
\subfloat[]{\includegraphics[width=0.48\textwidth]{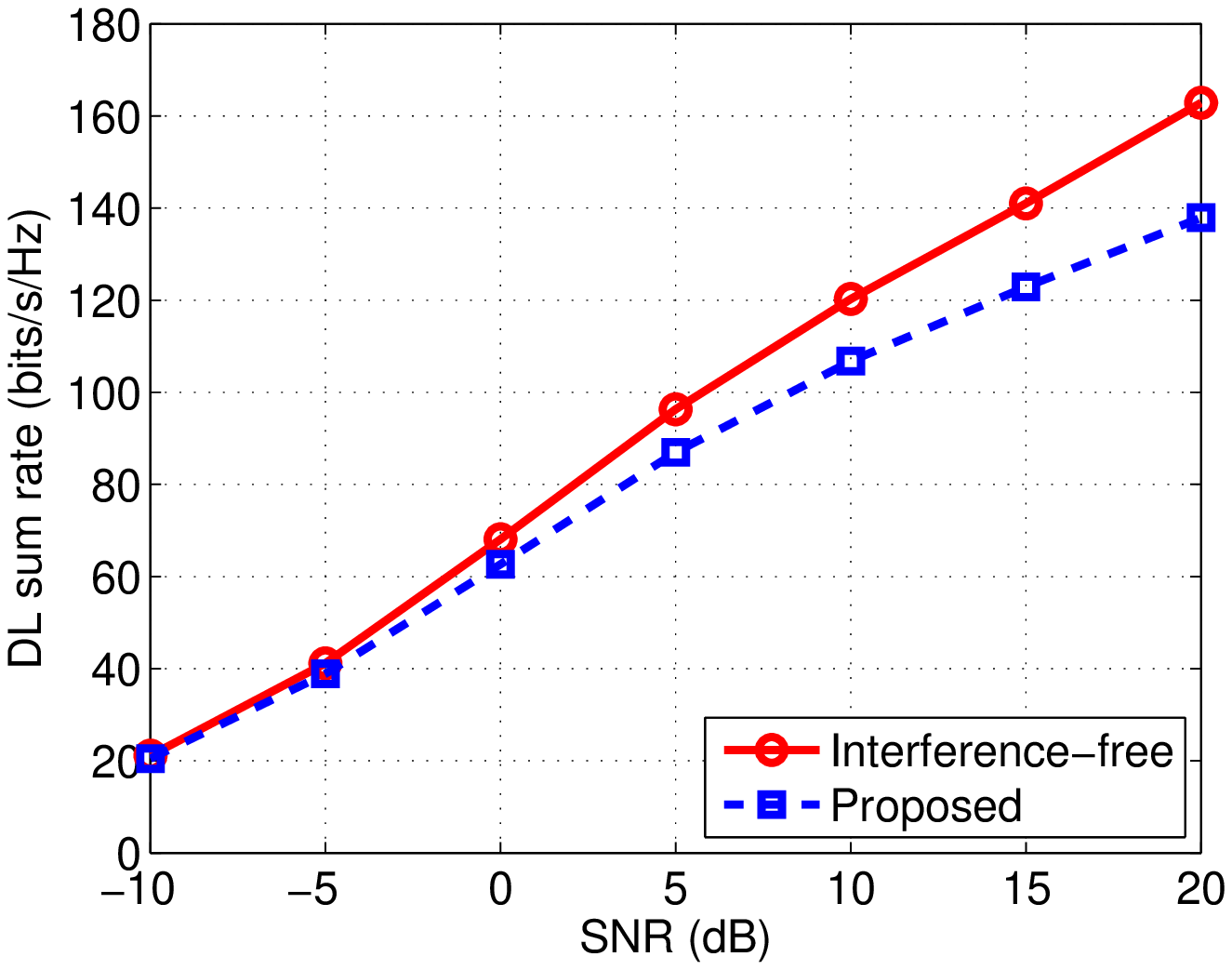}
\label{fig:bm_sche_carr_30g}}
\hfill
\subfloat[]{\includegraphics[width=0.48\textwidth]{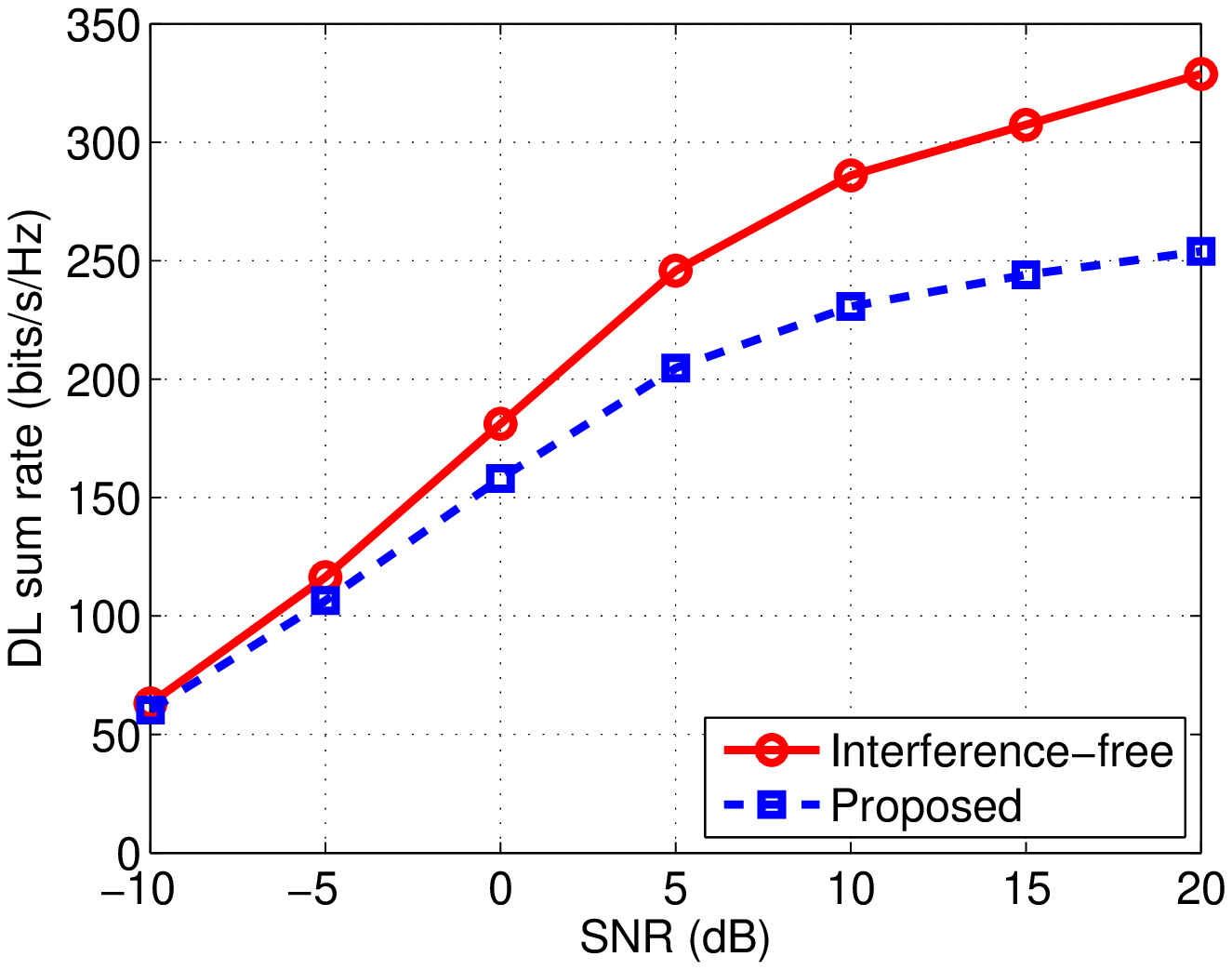}
\label{fig:bm_sche_carr_300g}}
\caption{Comparison of the DL sum rates of the interference-free case and the proposed beam scheduling algorithm. (a) 30 GHz; (b) 300 GHz.}
\label{fig:bm_sche_comp}
\end{figure*}

We first evaluate the performance of the proposed beam scheduling algorithm listed in \tabref{tab:GBSA}. As it is difficult to perform exhaustive search for the considered beam scheduling problem, an extreme case, namely, interference-free case, in which the inter-user interference is ``genie-aided'' eliminated, is adopted as the performance comparison benchmark. Note that the performance achieved by the optimal exhaustive search will lie between those of the interference-free case and the proposed algorithm. In \figref{fig:bm_sche_comp}, the achieved DL sum rates of the interference-free case and the proposed beam scheduling algorithm are presented. We can observe that the performance of the proposed beam scheduling algorithm can approach that of the interference-free case, especially in the low-to-medium SNR regime. In particular, for SNR=5 dB, the performance of the proposed algorithm can achieve at least 90\% and 83\% of the optimal exhaustive search at carrier frequencies of 30 GHz and 300 GHz, respectively. In the subsequent simulation, we will adopt the proposed beam scheduling algorithm to save the computational cost.

\begin{figure*}[!t]
\centering
\subfloat[]{\centering\includegraphics[width=0.48\textwidth]{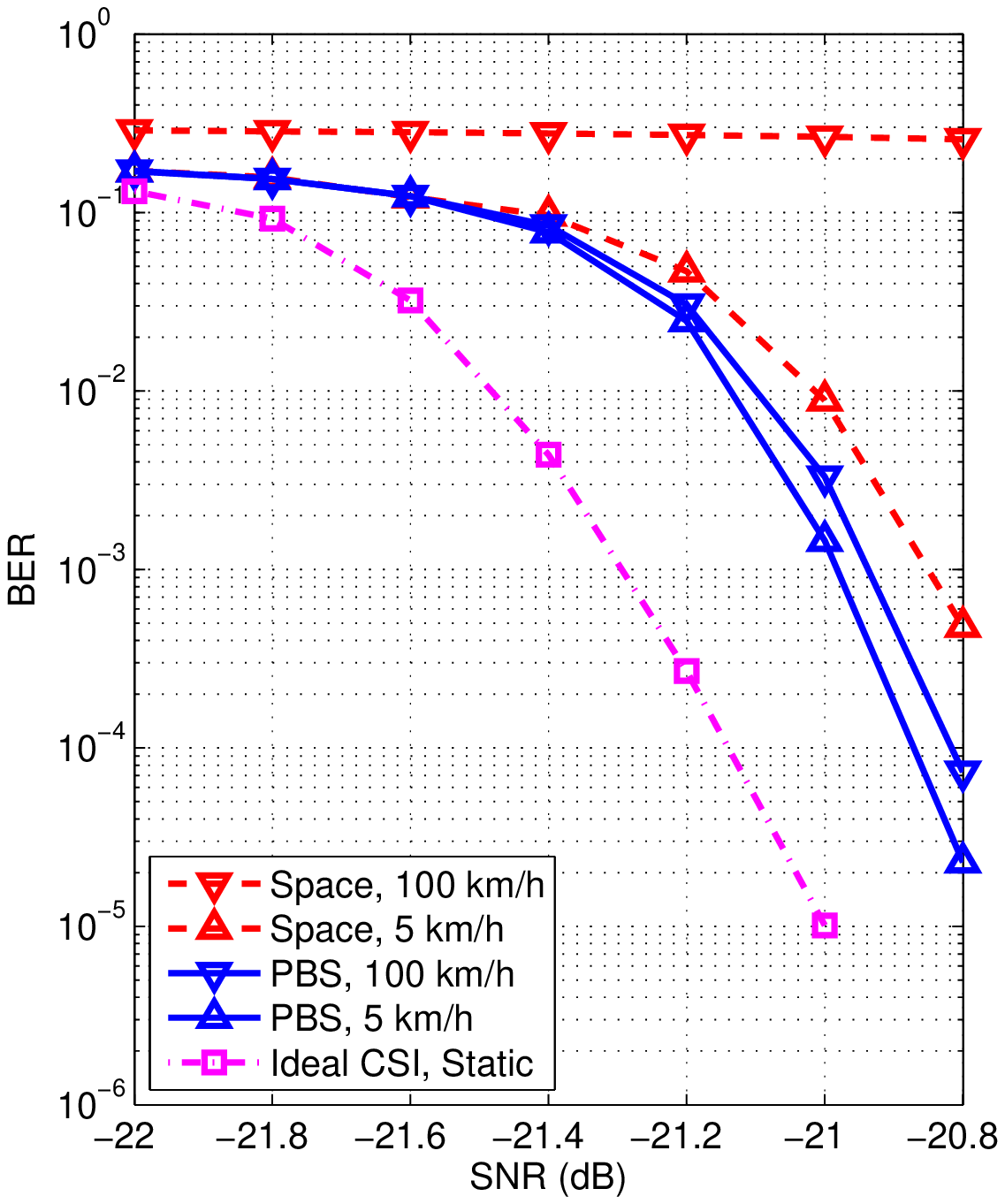}
\label{fig:ber_comp_carr_30g}}
\hfill
\subfloat[]{\centering\includegraphics[width=0.48\textwidth]{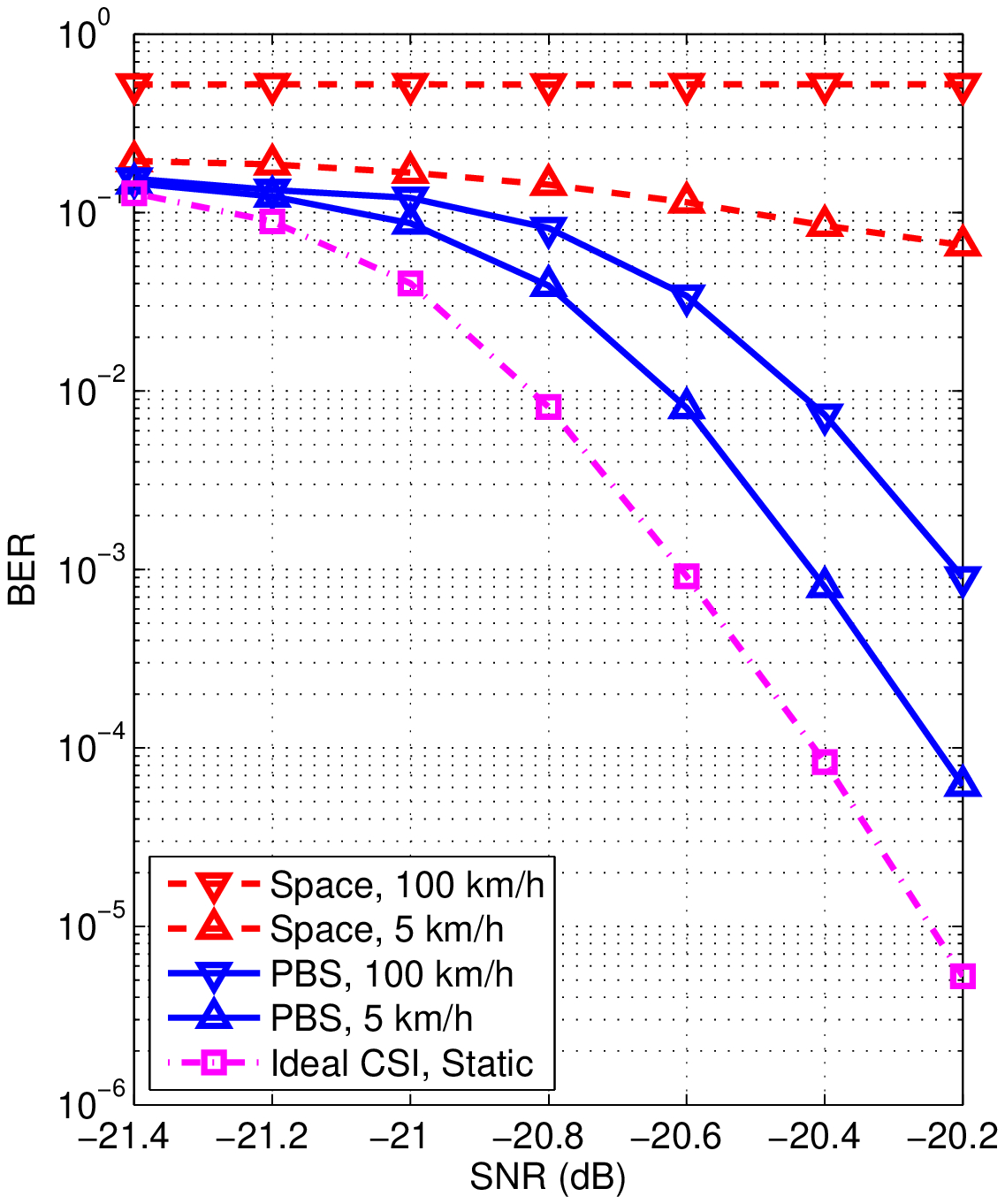}
\label{fig:ber_comp_carr_300g}}
\caption{Comparison of the BER performance with the proposed PBS and conventional space domain synchronization for 1/2-rate turbo-coded and QPSK mapped signals. The BER performance of the ideal case where the receivers have perfect instantaneous CSI for the static channels is also presented. (a) 30 GHz; (b) 300 GHz.}
\label{fig:ber_comp}
\end{figure*}

We then evaluate the performance of the proposed PBS. We focus on the bit-error rate (BER) performance of BDMA transmission for 1/2-rate turbo-coded quadrature phase-shift keying (QPSK) mapped signals and adopt the following simulation settings. Each transmission frame begins with one pilot OFDM symbol using the pilot design suggested in \cite{Sun15Beam}, followed by six data OFDM symbols. An iterative receiver as introduced in \cite{Wang06Low} is adopted. In \figref{fig:ber_comp}, the BER performance of the proposed PBS and conventional space domain synchronization under typical mobility scenarios is presented. The BER performance of the ideal case, where the receivers have perfect instantaneous CSI for static channels, is presented as the comparison benchmark. We can observe that the proposed PBS outperforms conventional space domain synchronization significantly in typical mobility scenarios at mmW/THz bands, which demonstrates the effectiveness of the proposed PBS.

\section{Conclusion}\label{sec:pbs_conc}

In this paper, we have proposed BDMA for mmW/THz massive MIMO transmission with per-beam  synchronization (PBS). We have first investigated the physically motivated beam domain channel model and shown that when both the numbers of antennas at BS and UTs tend to infinity, the beam domain channel fading in time and frequency disappears asymptotically. This channel property has then motivated us to propose PBS, where signal over each beam of the UTs is synchronized individually.
We have shown that both the effective channel delay and Doppler frequency spreads can be approximately reduced by a factor of the number of UT antennas in the large array regime with PBS compared with the conventional synchronization approaches, which effectively mitigates the severe Doppler effect in mmW/THz systems and leads to a significantly reduced CP overhead. We have further applied PBS to BDMA. We have investigated beam scheduling for both UL and DL BDMA and a greedy beam scheduling algorithm has been developed. The effectiveness of the PBS-based BDMA for mmW/THz massive MIMO-OFDM systems in typical mobility scenarios has been verified in the simulation.


\end{document}